\documentclass[prd,nofootinbib,floats,superscriptaddress,eqsecnum,tightenlines,preprintnumbers,11pt]{revtex4}

\usepackage{hyperref}
\usepackage{graphicx}
\usepackage{amsmath,amssymb,amsfonts,amsthm,latexsym}
\usepackage{marginnote}
\usepackage{xcolor}
\usepackage{physics}
\usepackage{nicefrac}

\usepackage{amsmath}%
\usepackage{amsfonts}%
\usepackage{amssymb}%
\usepackage{graphicx}

\usepackage{mathrsfs}
\usepackage{xspace}
\usepackage{mathtools, amssymb}

\def\be{\begin{equation}}
\def\ee{\end{equation}}
\def\beq{\begin{equation}}
\def\eeq{\end{equation}}
\newcommand{\bea}{\begin{eqnarray}}
\newcommand{\eea}{\end{eqnarray}}
\def\bi{\begin{itemize}}
\def\ei{\end{itemize}}
\def\ba{\begin{array}}
\def\ea{\end{array}}
\def\bfig{\begin{figure}}
\def\efig{\end{figure}}

\newcommand{\AR}{{\mathscr{A}}}  
\newcommand{\AF}{{\mathscr{B}}}  
\newcommand{\AM}{{\mathscr{M}}}  

\newcommand{\aaa}{{\alpha}}  
\newcommand{\bbb}{{\beta}}

\newcommand{\EK}{{\mathscr{E}}}
\newcommand{\GK}{{\mathscr{G}}}
\newcommand{\VK}{{\mathscr{V}}}
\newcommand{\FFd}{P}  % F Fdual
\newcommand{\Ga}{G} % Gauss constraint
\newcommand{\N}{\Xi}
\newcommand{\M}{\Lambda}
\newcommand{\PP}{\Upsilon}
\newcommand{\Lag}{ \mathscr{L}}
\newcommand{\Q}{L}
\begin{document}

\title{Degenerate Higher-Order Maxwell Theories in Flat Space-Time}

\author{Aimeric Coll\'{e}aux}
\affiliation{Astrophysics Research Center, The Open University of Israel, Ra’anana, Israel}
\author{David Langlois}
\affiliation{Universit\'e Paris Cit\'e, CNRS, Astroparticule et Cosmologie, F-75013 Paris, France}
\author{Karim Noui}
\affiliation{Laboratoire de Physique des deux Infinis IJCLab, Universit\'e Paris-Saclay, CNRS, France}

\date{\today}

\begin{abstract}
We consider, in Minkowski spacetime,  higher-order Maxwell Lagrangians with terms quadratic in the derivatives of the field strength tensor, and study their degrees of freedom. Using a 3+1 decomposition of these Lagrangians,  we extract the  kinetic matrix for the components of the electric field, corresponding to second time derivatives of the gauge field. If the kinetic matrix is invertible, the theory admits five degrees of freedom, namely the usual two polarisations of a  photon plus three  extra degrees of freedom which are shown to be Ostrogradski ghosts. We also classify the cases where the kinetic matrix is non-invertible and, using analogous simple models, we argue that, even though the degeneracy conditions reduce the number of degrees of freedom, it does not seem possible to fully eliminate all potential Ostrogradski ghosts. 
 \end{abstract}

\maketitle

%\tableofcontents

\section{Introduction}
In a recent article \cite{Colleaux:2023cqu}, we have classified a large family of higher-order Einstein-Maxwell theories in 4 dimensions whose action $S[g_{\mu\nu},A_\mu]$  couples a metric tensor $g_{\mu\nu}$ with a $U(1)$-gauge field $A_\mu$ in a non-minimal derivative way. More specifically,  the corresponding Lagrangians, which are invariant under spacetime diffeomorphisms, include terms linear in the curvature tensor $R_{\mu\nu\rho\sigma}$ and terms quadratic in the covariant derivatives of the field strength tensor $F_{\mu\nu}=\partial_\mu A_\nu - \partial_\nu A_\mu$. 
Examples of such theories can in principle be constructed from disformal transformations of the usual Einstein-Maxwell action \cite{DeFelice:2019hxb,Gumrukcuoglu:2019ebp}.

As these Lagrangians involve second derivatives of  $A_\mu$ which cannot be eliminated by integrations by part in general,  they could lead to potentially dangerous Ostrogradski ghosts. 
To avoid this problem, one needs to look for  degeneracy conditions, similar to those that have been introduced in DHOST (Degenerate Higher-Order Scalar-Tensor) theories \cite{Langlois:2015cwa,Langlois:2015skt,BenAchour:2016cay,BenAchour:2016fzp}, in order  to ensure that the theory does not propagate an extra degree of freedom even though the equations of motion are higher order. However, the problem of finding necessary and sufficient degeneracy conditions to get rid of these extra degrees of freedom is much more complex  here than in higher-derivative scalar-tensor theories for the reason that the gauge field $A_\mu$ is a vector. 
For instance, while higher-order scalar-tensor Lagrangians can lead to at most one extra degree of freedom, higher-order Einstein-Maxwell theories could contain up to three extra degrees of freedom, which would require more constraints  to eliminate all of them. 

In this article, we study the degeneracy conditions in the much simpler limit where the metric is  flat. Hence, higher-order Einstein-Maxwell theories reduce to higher-order Maxwell theories $S[A_\mu]$ which are Lorentz and $U(1)$-invariant. Particular examples of such theories have already being studied in different contexts such as the effective actions of quantum electrodynamics \cite{Lee:1989vh,Cangemi:1994by,Gusynin:1998bt,Gusynin:1995bc,Navarro-Salas:2020oew,Karbstein:2017pbf} and the Bopp-Podolsky generalized electrodynamics \cite{Bopp1, Podolsky:1942zz, Bufalo:2010sb, El-Bennich:2020aiq,Cuzinatto:2016kjk, Ji:2019phv}. 
Higher-order theories that  lead to second-order field equations  in Minkowski spacetime reduce to the usual Maxwell theory, as implied by the results  of \cite{Horndeski:1976gi}.
This result has been partially extended to arbitrary spacetime dimension in the more recent article \cite{Deffayet:2013tca}.
Here we go further by allowing equations of motion with order higher than 2 and argue that one cannot find higher-order Maxwell theories, at most quadratic in the derivatives of  $F_{\mu\nu}$, without ghost-like degrees of freedom. 

Our analysis proceed as follows. First, we decompose  the most general  Lagrangian in Minkowski with terms at most quadratic in $\partial_\alpha F_{\mu\nu}$ on an explicit basis of 17 elementary Lagrangians. We then use a 3+1 decomposition, compute the kinetic Lagrangian which is quadratic in the second time derivatives of
the gauge field, and  extract  the corresponding 3-dimensional kinetic matrix. We show that, when the kinetic matrix is non-degenerate, the theory admits 5 degrees of freedom: 2 of them are associated with the usual polarisations of the photon, while the remaining 3 degrees of freedom behave as Ostrogradski ghosts. 

It is interesting to note that two of these extra degrees of freedom are the Ostrogradski ghosts associated with the usual two polarisations  of $A_\mu$ while a kind of ``longitudinal'' mode ``arises" due to the presence of higher derivatives in the action. This is the reason why these extra degrees of freedom are sometimes interpreted as those of a massive photon (see for e.g. \cite{El-Bennich:2020aiq,Cuzinatto:2016kjk} and references therein). Therefore, requiring the degeneracy of the kinetic matrix is necessary to preserve the degrees of freedom of the Maxwell theory and evade the potential instabilities related to these extra ghosts\footnote{Although these instabilities could be fatal, in some situations higher order derivatives do not produce quantum instabilities, see for example \cite{Donoghue:2021eto}. From this perspective, the theories we obtain in the following could be worth investigating, despite their ghost-like degrees of freedom.}.
As expected, we show that imposing the degeneracy conditions  reduces the number of degrees of freedom.
However, we argue  (with the help of toy models and using explicit examples) that, even if the kinetic matrix is degenerate, any higher-order Maxwell theory seems to always propagate at least one Ostrogradski ghost. We illustrate this result with some examples.

\medskip

The paper is organised as follows. In the next section, we present the most general class of higher-order Maxwell Lagrangians that depend quadratically on the derivatives of the field strength tensor, in  Minkowski space. The field equations are derived and the Hamiltonian analysis of the non-degenerate theories is carried out. The case of degenerate higher-order Maxwell theories is investigated in section \ref{SecDHOMT}. We classify these theories into three classes according to the rank of their kinetic matrix. All  seven theories with rank zero (corresponding to quasi-linear theories) are constructed and it is shown that there are  3 degrees of freedom for generic coupling functions. When the rank is one, we find that imposing some conditions on the linear piece in the second derivative of the gauge potential reduces the number of degrees of freedom to $3$ as well. The resulting subclass of rank $1$ theories can be further restricted to yield at most two degrees of freedom, one of which being a ghost.
Finally, we conclude in section \ref{SecDisc} and add some technical details in appendices.

\section{Higher-Order Maxwell theories}\label{SecHOMT}
In this section, we  present the 4-dimensional Higher-Order Maxwell theories that we consider in this work  and carry out  a 3+1 decomposition of their action in order to perform their Hamiltonian analysis. This enables us not only to count the number of degrees of freedom but also to see whether Ostrogradski ghosts are propagating in these theories.

\subsection{Action and equations of motion}
In the present work, we consider actions  of the form 
 \begin{eqnarray}
S[A_\mu] = \int d^4 x \left(\AM + \AF^{\gamma\mu\nu,\delta\rho\sigma} \, \partial_\gamma F_{\mu\nu} \, \partial_\delta F_{\rho\sigma}  \right)
\; = \;  \int d^4 x  \, {\mathscr{L}} \, , \label{FlatAction}
\end{eqnarray} 
where $F_{\mu\nu}$ denotes the field strength tensor associated with the $U(1)$ gauge field $A_\mu$,
\begin{eqnarray}
 F_{\mu\nu}=\partial_\mu A_\mu - \partial_\nu A_\mu \,.
 \end{eqnarray}
While the six-index tensor $\AF$ depends on the field strength tensor, the scalar function $\AM$ is assumed to depend only on the two electromagnetic scalar invariants available in four dimensions, namely
\begin{eqnarray}
\label{F2P}
F^2 = F^{\mu\nu}F_{\mu\nu} \,, \qquad \FFd= {}^*\!F_{\mu\nu} F^{\mu\nu} = \frac{1}{2}\varepsilon_{\mu\nu\rho\sigma} F^{\rho\sigma} F^{\mu\nu} \, ,
\label{U1invariants}
\end{eqnarray}
where $*$ denotes the Hodge dual defined from the fully anti-symmetric  tensor $\varepsilon_{\mu\nu\rho\sigma}$ in four dimensions. 

The above family of theories thus contains the Maxwell action as a particular case with $\AM=F^2/4$ and $\AF=0$. The action (\ref{FlatAction}) can also be seen as the flat space limit of  higher-order Einstein-Maxwell theories studied  in our previous paper  \cite{Colleaux:2023cqu}. As a consequence, we know that the tensor $\AF$ can be decomposed, without loss of generality, into 17 elementary tensors, with coefficients that depend on the two scalars $F^2$ and $\FFd$, as summarised in Appendix \ref{App:class}.

The equations of motion derived from \eqref{FlatAction}  can be written in a  form similar to the vacuum Maxwell equations $\partial_\nu F^{\mu\nu}=0$, namely
\begin{eqnarray}
\label{modifiedMax}
\partial_\nu { H^{\mu\nu}} \; = \; 0 \, ,
\end{eqnarray}
with
\begin{eqnarray}
H^{\mu \nu} & = &  \frac{\partial {\mathscr{L}}}{\partial F_{\mu\nu}} - \partial_\alpha \left( \frac{\partial {\mathscr{L}}}{\partial \, \partial_\alpha F_{\mu\nu}}\right)   \\
& = & \frac{\partial \AM}{\partial F_{\mu\nu}} + 
 \frac{\partial \AF^{\gamma \alpha \beta,\xi \rho \sigma}}{\partial F_{\mu\nu}} \, \partial_\gamma F_{\alpha \beta} \, \partial_\xi F_{\rho \sigma} 
 -2 \partial_\alpha \left( \AF^{\alpha \mu\nu,\xi \rho \sigma} \partial_\xi F_{\rho \sigma}\right) .
\end{eqnarray}
Note that the ``modified'' Maxwell tensor $H^{\mu\nu}$ in general does not obey the Bianchi identity
\beq
\label{Bianchi}
\partial_{[\lambda}F_{\mu\nu]}=0\,,
\eeq
satisfied by the field strength. The  equations of motion (\ref{modifiedMax})  involve, in general, up to third derivatives of $F_{\mu\nu}$, hence  they are fourth order in the gauge field itself $A_{\mu}$.  

The action \eqref{FlatAction} yields higher order field equations and thus contains additional degrees of freedom for generic couplings compared to the usual Maxwell
theory which describes the dynamics of 2 degrees of freedom associated with the 2 polarisations of  the electromagnetic field. 
As these extra degrees of freedom are associated with higher derivatives in the action, one can expect them to be  Ostrogradski ghosts, as  we  confirm below.

\subsection{Hamiltonian decomposition of the action}
Similarly to the case of  higher order scalar-tensor theories (where the Lagrangian involves second derivatives of a scalar field), the Lagrangian must be degenerate in order to evade problematic Ostrogradski ghosts. A systematic way to identify the degrees of freedom consists in performing a Hamiltonian analysis of the action in order to extract the constraints and  see whether or not they are sufficient to eliminate the extra degrees of freedom.

\subsubsection{Equivalent formulation to eliminate higher derivatives}

As a first step, it is convenient to replace all second-order derivatives in the action by first-order derivatives, using  the equivalent action
\begin{eqnarray}
S_{\rm{eq}}[F_{\mu\nu},A_\mu,\lambda_{\mu\nu}] & = & \int d^4 x \sqrt{-g}  \left[ {\mathscr{L}} + \frac{1}{2} \lambda^{\mu\nu} \left( F_{\mu\nu}  - \partial_\mu A_\nu + \partial_\nu A_\mu \right) \right], \label{EqAction}
\end{eqnarray}
where the Lagrangian density ${\mathscr{L}}$ is the same as the one in \eqref{FlatAction}, 
\begin{eqnarray}
{\mathscr{L}}[F_{\mu\nu}]  = \AM + \AF^{\gamma\mu\nu,\delta\rho\sigma} \, \partial_\gamma F_{\mu\nu} \, \partial_\delta F_{\rho\sigma} \, ,
\end{eqnarray}
but  it is now viewed as a functional of  $F_{\mu\nu}$, which is  a dynamical anti-symmetric tensor  on its own, a priori unrelated to the gauge field $A_\mu$. 
It is only as the consequence of  the equation of motion obtained by varying the action
with respect to the antisymmetric tensor $\lambda_{\mu\nu}$ that  $F_{\mu\nu}$ turns out to be, on-shell, the strength field associated with $A_\mu$. Proceeding in this way enables us to recast higher derivatives in the action  as first order derivatives and then  use a standard  Hamiltonian analysis.

In this formulation, the equations of motion are obtained from the variation of  the action with respect to $F_{\mu\nu}$,  $A_\mu$ and $\lambda^{\mu\nu}$, yielding respectively 
\begin{eqnarray}
\frac{\partial {\mathscr{L}}}{\partial F_{\mu\nu}} - \partial_\alpha \left( \frac{\partial {\mathscr{L}}}{\partial \partial_\alpha F_{\mu\nu}}\right) +\frac{1}{2} \lambda^{\mu\nu}= 0 \, , \qquad
\partial_{\nu} \lambda^{\mu\nu} = 0 \, , \qquad F_{\mu\nu} = \partial_{\mu}A_{\nu} - \partial_\nu A_\mu \, .
\end{eqnarray}
Taking the divergence of the first equation above enables us to eliminate the new variable $\lambda_{\mu\nu}$ {(which is given by $\lambda= -2H$)} and thus recover the equation \eqref{modifiedMax}, which confirms
the equivalence between  the actions \eqref{FlatAction} and \eqref{EqAction}.

\subsubsection{3+1 decomposition of spacetime}
Working in Minkowski coordinates, where the metric reads $\eta_{\mu\nu}=\text{diag}(-1,+1,+1,+1)$, it is convenient, for the 3+1 decomposition of the action, 
to distinguish  spatial indices with  latin letters $(i,j,k,\cdots)$ from  spacetime indices denoted by greek letters $(\mu,\nu,\rho,\cdots)$.

We first decompose the gauge field into  $A_\mu=(A_0,A_i)$ and the field strength $F_{\mu\nu}$ into  its electric and magnetic components,
\begin{eqnarray}
E_i \equiv F_{0i}=\dot A_i-\partial_i A_0 \; , \qquad B_i \equiv \frac{1}{2} \varepsilon_{ijk} F^{jk}=\varepsilon_{ijk}\partial^j A^k \, ,
\end{eqnarray}
where $ \varepsilon_{ijk}$ is the fully antisymmetric 3-dimensional symbol.
Similarly,  the antisymmetric tensor $\lambda_{\mu\nu}$ introduced in the action \eqref{EqAction} can be decomposed into two spatial vectors, 
\begin{eqnarray}
\pi_i = \lambda_{0i} \, , \qquad \lambda_i =  \frac{1}{2} \varepsilon_{ijk} \lambda^{jk} \, .
\end{eqnarray}

From the definitions of $E$ and $B$, one can immediately write the various components of the higher-derivative tensors $\partial_\alpha F_{\mu\nu}$ in terms of the electric and magnetic fields, obtaining in particular
\begin{eqnarray}
\partial_0 F_{0i}= \dot E_i \, , \qquad 
\partial_j F_{0i} = \partial_j E_i \, , \qquad \partial_k F_{ij} = \varepsilon_{ijm} \partial_k B^m \,.
\end{eqnarray}
For the time derivative of the spatial components or, equivalently, of the magnetic field, it is convenient to use the  Bianchi identities \eqref{Bianchi} to transform them into space derivatives of the electric field so that 
\beq
\partial_0 F_{ij} = \partial_i E_j - \partial_j E_i\,.
\eeq
As a consequence, no time derivatives of $B$ remains in the action.

Substituting all the above formulas into the action \eqref{EqAction}, one obtains its 3+1 decomposition which, after a few integrations by parts, takes the form
\begin{eqnarray}
\label{actioneqEE}
S_{\rm{eq}}  =  \int dt \int d^3 x  \, \left( 
\frac{1}{2}\EK^{ij} \dot E_i \dot E_j +\GK^i \dot E_i  + 
\pi^i \dot A_i  -  \pi^i E_i  + A_0 \Ga + \lambda_i \chi^i  - \VK  \right),
\end{eqnarray}
where the symbols $\EK$, $\GK$  and $\VK$ are defined as follows.
The kinetic matrix $\EK$ and the vector field $\GK$ are given by 
\begin{eqnarray}
\label{Gigen}
\EK^{ij}= 8  \, \AF^{00i,00j} \, , \qquad \GK_i  = \Sigma^p_{ij}  \, \partial_p E^j + \Upsilon^p_{ij}  \, \partial_p B^j \, ,
\end{eqnarray}
with
\begin{eqnarray}
\Sigma^{pij} = -8 (\AF^{00i,0pj} + \AF^{00i,p0j}) \, , \qquad 
\Upsilon^{pij} = 4 \varepsilon^j{}_{kl} \, \AF^{00i,pkl} \, .
\end{eqnarray}
The explicit form of the potential $\VK$ is
\begin{eqnarray}
\VK = \VK_1^{ijkl } \, \partial_i E_j \, \partial_k E_l  + \VK_2^{ijkl }\, \partial_i B_j \, \partial_k B_l + \VK_3^{ijkl } \, \partial_i E_j \, \partial_k B_l  - \AM \, ,
\end{eqnarray}
where $\VK_n$ are the following combinations  of the components of $\AF$,
\begin{eqnarray}
 \VK_1^{ijkl } &=& -4 (\AF^{0ij,0kl} + \AF^{i0j,k0l} + \AF^{i0j,0kl}+ \AF^{0ij,k0l} ) \, , \\
 \VK_2^{ijkl } &= &-  \varepsilon^j{}_{rs} \,  \varepsilon^l{}_{mn} \, \AF^{irs,kmn} \, , \\
 \VK_3^{ijkl } &= &  4 \,  \varepsilon^l{}_{nm} ( \AF^{i0j,knm} +\AF^{0ij,knm}  )\, .
\end{eqnarray}
Finally, we see that $A_0$ and $\lambda_i$ are  Lagrange multipliers which enforce respectively the constraints\footnote{We are using the standard notation $\simeq$ for the 
weak equality, which is defined as an equality up to constraints.}
\begin{eqnarray}
\label{GchiBflat}
\Ga = \partial_i \pi^i \, \simeq 0 \,  , \qquad \chi_i = B_i - \varepsilon_{ijk}\partial^j A^k \ \simeq 0 \, .
\end{eqnarray}
We recognise the Gauss contraint $\Ga$ which generates the $U(1)$ gauge transformations.

\subsubsection{Phase space and constraint analysis}
From the 3+1 decomposition, we see that we can parametrise the 
phase space with the following 3 pairs of conjugate variables,
\begin{eqnarray}
\{ E_i (x) , \pi_E^j(y)\} \, = \, \{ B_i(x) , \pi_B^j(y)\} \, = \, \{A_i(x),\pi^j(y)\} \, = \, \delta (x-y) \, \delta_i^j \, ,
\end{eqnarray}
as it is clear from (\ref{actioneqEE}) that $\pi^i$ and $A_i$ are canonically conjugate variables. Since $A_0$ and $\lambda_i$ are Lagrange multipliers, as noticed above,   it is not necessary to introduce their conjugate momenta. 

There is no time derivative of the magnetic field in the action, hence we  get, in addition to the constraints \eqref{GchiBflat}, three additional  primary
constraints,
\begin{eqnarray}
\label{piB}
\psi^i \; = \; \pi_{B}^i \; \simeq \; 0 \, .
\end{eqnarray}
Finally, the expressions of the conjugate momenta of the electric field components are given by
\begin{eqnarray}
\label{piE}
\pi_E^i \; = \; \EK^{ij} \dot E_j + \GK^i \, .
\end{eqnarray}
These relations can be inverted only if the matrix of coefficients $\EK^{ij}$ is invertible, i.e. of rank $3$. We discuss this case in the next subsection. Otherwise,  we obtain additional  primary constraints, whose number depends on the rank, $\text{rk}(\EK)$, of the three-dimensional kinetic matric $\EK$, whose expression is given in Appendix \ref{App}. These degenerate situations  will be discussed in the next section.

\subsubsection{Non-degenerate theories}
When $\text{rk}(\EK)=3$, the matrix $ \EK^{ij} $ is invertible, and  we can  then express $\dot E_i$ in terms of the momenta by inverting (\ref{piE}). In this case, the total Hamiltonian is of the form
\begin{eqnarray}
H_{\rm{tot}} = \int d^3x \, \left(
\frac{1}{2} \EK^{ {-1}}_{ij} (\pi_E^i - \GK^i )(\pi_E^j -  \GK^j) +  \VK  
+  \pi^i E_i 
 -A_0 \Ga - \lambda^i \chi_i + \mu^i \psi_i 
\right) \, ,
\end{eqnarray}
where $\EK^{-1}$ denotes the inverse of $\EK$ and  we have introduced the new Lagrange multipliers $\mu^i$ to enforce the primary constraints (\ref{piB}).

Let us discuss the different constraints. We know that the Gauss constraint  is first class as it generates the $U(1)$ gauge symmetry. 
The other constraints $\chi_i$ and $\psi_i$  satisfy the Poisson relations
\begin{eqnarray}
\{ \chi_i(x) \, , \psi^j(y) \} \; =\{ B_i(x) - \varepsilon_{ilm}\partial^l A^m \, , \pi_B^j(y) \} \; = \;\delta(x-y) \,  \delta_i^j \, ,
\end{eqnarray}
which shows that they form a subset of second class constraints. As a consequence, requiring the time invariance of these constraints does not lead to new constraints, but rather enables to fix the Lagrange multipliers $\lambda^i$ and $\mu^i$ in terms of the other phase space variables. In practice, this implies that $\chi_i$ and $\psi^j$ can be set to strongly vanish, thus eliminating the pairs of variables $(B_i, \pi_B^i)$, which amounts to explicitly solve $B_i$ in terms of $A_i$ and to fix $\pi_B^i=0$.

In conclusion, the canonical analysis is completed with 6 second class constraints and 1 first class constraint, which results in $(18-6-1\times 2)/2=5$ degrees of freedom. In addition to the usual two electromagnetic field polarisations, we now have three extra 
degrees of freedom. The latter can be seen as Ostrogradski ghosts since the Hamiltonian is linear in the three momenta $\pi_i$, thus unbounded from below and from above in the three directions spanned  by the $\pi_i$. 

Interestingly, the fact that we have second time derivatives in the action leads not only to the emergence of the two expected  Ostrogradski ghosts associated with each of the polarisations, but also to a ghostly longitudinal mode. Somehow, the gauge degree of freedom acquires an Ostrogradski ghost which becomes physical.
{Notice that this has been advocated as a mechanism to generate a massive gauge field from the higher order self-interactions of a massless one in the context of the Bopp-Podolsky theory, see for instance \cite{El-Bennich:2020aiq,Ji:2019phv}.}

In the next section, we explore the possibilities to eliminate all or  some of the extra degrees of freedom by assuming the degeneracy of the kinetic matrix, i.e. 
 the non-invertibility of the  matrix $ \EK$, leading to additional constraints in the phase space.

\section{Degenerate Higher-Order Maxwell theories}\label{SecDHOMT}
In this section, we study degenerate higher-order Maxwell theories and  explore how many additional  constraints can be obtained, thus eliminating some or all the extra degrees of freedom identified in the previous section. 

\subsection{Quasi-linear theories: $\text{rk}(\EK)=0$. }
We first consider the case where the kinetic matrix vanishes, i.e.
\begin{eqnarray}
\EK^{ij} =0 \label{E=0} \, .
\end{eqnarray}
When this condition is satisfied, the action is said to be quasi-linear, i.e. at most linear in the second derivative of the electromagnetic potential (which is similar to General Relativity or more generally Lovelock-Lanczos gravity in higher dimensions). 

\subsubsection{General Lagrangian for quasi-linear theories}
Requiring \eqref{E=0} drastically restricts the number of independent terms in the action  \eqref{FlatAction}, as it can be seen from the explicit expression of the kinetic matrix written in Appendix \ref{App}. Instead of 17 independent Lagrangian terms quadratic in the derivatives of the field strength, only seven independent combinations  are allowed,
\begin{eqnarray}
\label{actiontk0}
S[A_\mu]  =  \int d^4 x  \, \big( \AM +  \sum_{p=1}^7 \alpha_p \, \mathscr{L}_p \big) \,,
\end{eqnarray}
where $\alpha_p$ are seven arbitrary functions of the two $U(1)$-invariants \eqref{U1invariants}. The seven independent combinations  $\mathscr{L}_p$ can be chosen as
\begin{eqnarray}
\mathscr{L}_{1}  &= &\mathcal{F}_{3}-\mathcal{F}_{5} =  
F_2^{\mu\nu} \partial_{[\sigma|}F_{\mu}{}^\sigma \, \partial_{|\gamma]}F_\nu{}^\gamma\,, \nonumber \\
\mathscr{L}_{2} & = & \mathcal{F}_{6}+\mathcal{F}_{8}= F^{\mu\nu} F^{\rho\sigma} \partial_{[\nu|}F_{\rho\mu} \, \partial_{|\gamma]}F_\sigma{}^\gamma      \,,\nonumber \\
\mathscr{L}_{3} &= &\mathcal{F}_{7}-\mathcal{F}_{8}= \frac{1}{2} F^{\mu\nu}  \partial_{[\mu} F^2 \, \partial_{\gamma]} F_\nu{}^\gamma   \,,  \label{QLT2} \\
\mathscr{L}_{4}  & =  &\mathcal{F}_{10}+\mathcal{F}_{11}= F^{\mu\nu} F_3^{\rho\sigma}  \partial_{[\nu|}F_{\sigma}{}^{\gamma} \, \partial_{|\gamma]}F_{\mu \rho}   \,, \nonumber \\ 
\mathscr{L}_{5} &= & \frac{\FFd^2}{16} \mathcal{F}_{1}- \frac{F^2}{2}\mathcal{F}_{8} -2\mathcal{F}_{11} +\mathcal{F}_{13}    \,, \nonumber \\ 
\mathscr{L}_{6} &= & \mathcal{F}_{14}-\mathcal{F}_{15}= \frac{1}{2}F^{\mu\nu}F_2^{\rho\sigma}   \partial_{[\mu|} F^2  \, \partial_{|\sigma]} F_{\nu \rho}   \,, \nonumber \\ 
\mathscr{L}_{7} &= & \mathcal{F}_{17}= F^{\mu\nu}F_2^{\rho\sigma}F_2^{\gamma\delta} \partial_{[\nu|} F_{\mu\rho} \, \partial_{|\delta]} F_{\sigma \gamma}  \,,\nonumber
 \end{eqnarray}
 where the basis Lagrangians $ \mathcal{F}_{a}$ are defined in equations (\ref{F_1}-\ref{F_18}) of  the appendix \ref{App:class}.

 Interestingly, all the above combinations can be rewritten in the form
 \begin{eqnarray}
 \label{antisymcov}
\mathscr{L}_p =  \AF^{\gamma\mu\nu,\delta\rho\sigma}_p \partial_{[\gamma\vert} F_{\mu\nu}\, \partial_{\vert\delta]} F_{\rho\sigma}  = 
 \AF^{[\gamma\vert \mu\nu,\vert \delta]\rho\sigma}_p \partial_{\gamma} F_{\mu\nu}\, \partial_{\delta} F_{\rho\sigma} \, ,
 \end{eqnarray}
with different tensors $\AF_p$.
To obtain this manifestly  antisymmetric expression for $\mathscr{L}_{3}$, $\mathscr{L}_{6}$ and $\mathscr{L}_{7}$, we used  the Bianchi identity (\ref{Bianchi}), whereas  for $\mathscr{L}_{5}$, we needed a dimensionally dependent identity (see \cite{Colleaux:2023cqu} for details) to rewrite it in the form
\begin{eqnarray}
\mathscr{L}_{5}=  F^{\mu\nu}F_3^{\rho\sigma} \left(\partial_{[\nu|}F_{\sigma \gamma} \, \partial_{|\rho]}F_{\mu}{}^{\gamma}+\partial_{[\nu|}F_{\mu \rho} \, \partial_{|\gamma]}F_{\sigma}{}^{\gamma}-\frac{1}{2}\partial_{[\nu|}F_{\mu}{}^\gamma \, \partial_{|\gamma]}F_{\rho\sigma} - \partial_\mu F_\rho{}^\gamma \,  \partial_\nu F_{\sigma \gamma} \right).
\end{eqnarray}
In the general expression (\ref{antisymcov}), one sees that  the antisymmetry property applies to the two indices of the derivatives, which makes transparent the fact that these Lagrangians cannot give 
 terms quadratic in $\dot F_{0i}$, as required for quasi-linear theories.
 
 \subsubsection{Constraint analysis} 
We now apply the Hamiltonian analysis of the previous section to our general quasi-linear action (\ref{actiontk0}). 
Because of the  condition \eqref{E=0}, the theory admits, in addition to \eqref{GchiBflat} and \eqref{piB},  3 additional primary constraints given by
\begin{eqnarray}
\phi_i \; = \; \pi_{Ei} - \GK_i \; \simeq \; 0 \, ,
\end{eqnarray}
as a consequence of \eqref{piE}.
The total Hamiltonian is now given by
\begin{eqnarray}
H_{\rm tot} =  \int d^3x \, \left( \VK -A_0 \Ga - \lambda_i \chi^i + \mu_i \psi^i + \nu_i \phi^i   + E^i \pi_i \right) \, ,
\end{eqnarray}
where we have introduced the  Lagrange multipliers $\nu_i$ to  enforce the new constraints $\phi_i \simeq 0$. 
The constraints $\chi_i$ and $\psi_i$ still form a subset of second class constraints and requiring their stability under time evolution does not lead to new constraints but, instead,  fixes the Lagrange multipliers $\lambda_i$ and $\mu_i$ in terms of the phase-space variables. 

Studying the time evolution of the constraints $\phi_i$ is more subtle. In order to compute $\dot \phi_i$, we  recall that $\GK_i$ is of the form \eqref{Gigen}
where ${\Sigma}_{ij}^{p}$ and ${\Upsilon}_{ij}^{p}$ depend on the components of $E$ (and also $B$) but not on its derivatives. Then a direct calculation shows that
\begin{eqnarray}
\dot \phi_i(x) = \{ \phi_i(x), H_{\rm tot} \} = \Delta_{ij} \nu^j(x) - \pi_i(x) + \{ \phi_i(x) , \tilde H\} \, ,
\end{eqnarray} 
where we have introduced
\beq
\tilde H=  \int d^3x \, \left( \VK -A_0 \Ga - \lambda_i \chi^i + \mu_i \psi^i\right) 
\eeq
and 
where $\Delta_{ij}$ has to be understood, in general, as a differential operator acting on any vector field $\nu^i(x)$ as follows,
\begin{eqnarray}
\Delta_{ij} \nu^j\; = \; \{ \phi_i(x), \phi[\nu] \} \quad \text{where} \quad
\phi[\nu] = \int d^3x \, \phi_j(x) \, \nu^j(x) \, .
\end{eqnarray}
Since ${\GK}_i$ is of the form \eqref{Gigen}, we see that the operator $\Delta_{ij}$ can be decomposed into two parts, according to
\begin{eqnarray}
\Delta_{ij} \; = \;  \Delta_{ij}^{0} - S_{ij}^p \, \partial_p \,,
\end{eqnarray}
with
\begin{eqnarray}
 \Delta_{ij}^{0} =  \left(\frac{\partial\Upsilon_{jk}^p}{\partial E^i}- \frac{\partial\Upsilon_{ik}^p}{\partial E^j} \right) \partial_p B^k 
 + \left( \frac{\partial \Sigma_{jk}^{p} }{\partial E^i} - \frac{\partial \Sigma_{ik}^{p} }{\partial E^j}  \right) \partial_p E^k - \partial_p \Sigma^p_{ji} \, , 
 \qquad 
 S_{ij}^p =  \Sigma_{ij}^{p} + \Sigma_{ji}^{p} \,  .
 \end{eqnarray}
Moreover, the quasi-linear theories
 have the nice property that $S_{ij}^p$ vanishes identically. This is a direct consequence of the antisymmetry between the  derivatives  (of the field strengths) in the Lagrangian  \eqref{antisymcov}. Indeed, $\Sigma^p_{ij}$ can be shown to be given by
\begin{eqnarray}
{\Sigma}^{p, ij} =  - \sum_{p=1}^5 \alpha_p \left( \AF^{0 0 i ,p 0 j}_p - \AF^{00 j , p0i}_p +\AF^{p 0 j ,0 0 i}_p - \AF^{p 0 i ,0 0 j}_p   \right) \, ,
 \end{eqnarray}
 which is antisymmetric in the indices $(i,j)$, so that $S_{ij}^p=0$. Hence, requiring the time invariance of $\phi_i$ leads to  algebraic relations between the Lagrange multipliers
 $\nu_i$.  
 To go further in the  analysis, we need to study the properties of the matrix $\Delta_{ij}=\Delta^0_{ij}$. 
 
 If $\Delta^0_{ij}$ vanishes
identically, then we get three additional (secondary) constraints
\begin{eqnarray}
\xi_i \; = \;  \pi_i - \{ \phi_i , \tilde H \} \; \simeq \; 0 \, ,
\end{eqnarray}
which enable us to solve the momenta $\pi_i$ in terms of the other phase space variables. Thus, these momenta disappear from
the Hamiltonian which implies that there is no Ostrogradski instability. In fact, it is easy to see that the combined  maximal degeneracy conditions
\begin{eqnarray}
\EK^{ij}=0 \, , \qquad
\Delta_{ij}=0 \, ,
\label{fulldegeneracy}
\end{eqnarray} 
are equivalent to the condition that the equations of motion for the gauge field  $A^i$ are (at most) second order. Indeed, the equations of motion for $A_i$ take
 the form
\begin{eqnarray}
- \EK^{ij} \dddot{E}_j + \left( \frac{\partial \EK^{kj}}{\partial E_i} -2  \frac{\partial \EK^{ik}}{\partial E_j}  - \frac{\partial \EK^{ij}}{\partial E_k}\right)   \dot E_j \ddot E_k  -2 \frac{\partial \EK^{ik}}{\partial B_j} \dot B_j  \ddot E_k + \Delta^{ij} \ddot E_j + R^i = 0 \, , 
\end{eqnarray}
where
\begin{eqnarray}
E_i = \dot A_i - \partial_i A_0 \,,
\end{eqnarray}
and the ``rest'' $R^i$ is at most second order in $A_i$. Therefore,  the conditions \eqref{fulldegeneracy} imply that all  second and third time derivatives of $E^i$ disappear in the equations of motion, leaving only terms at most second order in time derivatives of $A^i$.   As  originally shown in  \cite{Horndeski:1976gi} and more recently in \cite{Deffayet:2013tca}, these theories correspond to non-linear electrodynamics where 
$\AF^{\gamma\mu\nu,\delta\rho\sigma} =0$.

This result is of course consistent with the absence of extra degrees of freedom:
the extra constraints $\phi_i \simeq 0$ and $\xi_i \simeq 0$ enable us to kill (at least) 3 additional degrees of freedom and then the theory contains (at most) 2 degrees of freedom, the expected two polarisations. If some of these extra constraints are first class or if they lead to new tertiary constraints, one could even get 1 or 0 degree freedom.

\medskip

If $\Delta_{ij}$ does not vanish then the analysis becomes more involved and it is difficult to see how  one could get enough constraints  to eliminate all the extra degrees of freedom. As $\Delta_{ij}$ is antisymmetric and odd-dimensional, it admits one null direction, which we denote  $\kappa_i$. This
implies that the time invariance of $\phi_i$ leads to a secondary constraint given by
\begin{eqnarray}
\label{new_constraint}
\kappa^i \dot \phi_i \; = \;  \kappa^i \left(- \pi_i(x) + \{ \phi_i(x) , \tilde H\}\right)  \; \simeq 0 \, .
\end{eqnarray}
This constraint enables us to solve $\kappa^i \pi_i$ in terms of the (regular) phase space variables. The next step would be to impose the time invariance of this new constraint and  see whether it provides new  constraints, and so on, until no new constraint appears.

If the complete Hamiltonian analysis does not provide any other constraint after (\ref{new_constraint}),  we end up with 4 second class constraints, which leaves 3 degrees of freedom. In practice, the rest of the analysis can become quite involved since  $\tilde H$   is already complicated and  the expressions of  the constraints themselves become more and more burdensome. This is why we have not tried to pursue the analysis further. However, we do not expect to find Lagrangians where all the extra degrees of freedom can be eliminated. Our expectation relies on an analogous toy model, studied below, which captures the essential features (although not all) of the theories characterised by $\EK^{ij}=0$ and
$\Delta_{ij}\neq 0$.

\subsubsection{A toy model for quasi-linear theories}

We now  introduce a toy model that retains the main features of the theories discussed above, but without the presence of cumbersome spatial derivatives.

This analogous model describes the interactions between three point particle-like degrees of freedom, denoted $A_i$ for $i \in \{1,2,3\}$, and 
is defined by the action
\begin{eqnarray}
S_{\rm{toy}}[A_i] = \int dt \, \left( C_i \ddot A_i  - V \right) \, ,
\end{eqnarray}
where $C_i$ and $V$ are generic functions of the variables $(\dot A_i, A_i)$.  It is equivalent to
\begin{eqnarray}
\label{linearterm}
S_{\rm{toy,eq}}[E_i,A_i,\pi_i] = \int dt \, \left( C_i \dot E_i  - V+ \pi_i(\dot A_i - E_i) \right) \, .
\end{eqnarray}
This model is similar to the action (\ref{actioneqEE}) with $\EK^{ij}=0$ (in the quasi-linear case) but  it is much simpler to analyse since it contains no  spatial derivative, no  magnetic field and is not $U(1)$-invariant. 

To further  simplify the action \eqref{linearterm}, one can add the  boundary term
\begin{eqnarray}
\int dt \, \frac{d J}{dt} \, \qquad \text{such that} \qquad \frac{\partial J}{\partial E_3} = -C_3 \, ,  
\end{eqnarray}
where $J$ is a function of $(E_i,A_i)$. In this way, we can formally eliminate $C_3$, up to a redefinition of  $C_1$, $C_2$ and $V$. Hence, without loss of generality, we can replace the action \eqref{linearterm} by the following one,
\begin{eqnarray}
\label{linearterm2}
S_{\rm{toy}}[E_i,A_i,\pi_i] = \int dt \, \left( C_1 \dot E_1 +  C_2 \dot E_2  - V + \pi_1(\dot A_1 - E_1) + \pi_2(\dot A_2 - E_2) \right) \, ,
\end{eqnarray}
where now $C_i$ and $V$ are generic functions of the variables $(E_1,E_2)$, $(A_1,A_2,A_3)$ and $\dot A_3$. 

As in the full theory, we get two primary constraints
associated with the momenta $\pi_{E_1}$ and $\pi_{E_2}$, canonical conjugates of $E_1$ and $E_2$ respectively:
\begin{eqnarray}
\phi_1 = \pi_{E_1} - C_1 \simeq 0 \, , \qquad \phi_2 = \pi_{E_2} - C_2 \simeq 0 \,.
\end{eqnarray}
The conjugate momentum of $A_3$ is given by
\begin{eqnarray}
\label{pi3}
\pi_3 = \frac{\partial C_1}{\partial \dot A_3} \dot E_1 + \frac{\partial C_2}{\partial \dot A_3} \dot E_2 + \frac{\partial V}{\partial \dot A_3}  \, .
\end{eqnarray} 
The condition $\Delta_{ij} \neq 0$ translates here  into 
\begin{eqnarray}
\label{Delta12}
\Delta_{12} = \{ \phi_1, \phi_2\} = \frac{\partial C_2}{\partial E_1} - \frac{\partial C_1}{\partial E_2} \neq 0 \, .
\end{eqnarray}
Therefore, requiring the  time invariance of the constraints $\phi_i$ cannot lead to secondary constraints. 

Hence, the only way  to get another constraint  is to require that \eqref{pi3}
is  also a primary constraint, which means that $\pi_3 $ should not depend on velocities. 
This is the case if the $C_i$ do not depend on $\dot A_3$ and if  $V$ is linear in $\dot A_3$.
This corresponds to an initial action of the form 
\begin{eqnarray}
\label{Stoy2}
S_{\rm{toy}} = \int dt \;  \left( C_1 \, \ddot A_1 \; + \; C_2 \, \ddot A_2 \, + C_3 \, \dot A_3 \; - W \right) \, ,
\end{eqnarray} 
or, equivalently,
\begin{eqnarray}
S_{\rm{toy,eq}} = \int dt \;  \left( C_1 \, \dot E_1 \; + \; C_2 \, \dot E_2 \, + C_3 \, \dot A_3 \; - W + \pi_1(\dot A_1 - E_1) + \pi_2(\dot A_2 - E_2) \right) \, ,
\end{eqnarray} 
where  $C_i$ and $W$ do not depend on $\dot A_3$.  This leads to a new primary constraint $ \pi_3 - C_3 \simeq 0$, 
but it turns out that this eliminates a regular degree of freedom, instead of one of the extra degrees of freedom. 
Indeed, the change of variable
\begin{eqnarray}
 A_3 \longrightarrow \tilde A_3(A_3,\dot A_1,\dot A_2,A_1,A_2) \quad \text{such that} \quad \frac{\partial  \tilde A_3}{\partial A_3} = C_3   \, ,
\end{eqnarray}
transforms the action \eqref{Stoy2} into
\begin{eqnarray}
S_{\rm{toy}} = \int dt \;  \left( \tilde C_1 \, \ddot A_1 \; + \;  \tilde C_2 \, \ddot A_2 \, - \tilde W  \right)\, ,
\end{eqnarray}
up to boundary terms. There is no need to go further to understand that the theory propagates Ostrogradski ghosts \cite{Motohashi:2014opa}.
We expect a similar  scenario to happen when considering the full theory.

As a conclusion, this toy model strongly  suggests that, when $\text{rk}({\EK}) =0$,  the only way to avoid Ostrogradki ghosts  is to require second order equations of motion, i.e.
$\Delta_{ij}=0$. In this case,  there are no higher order terms in the quasi-linear Lagrangian \eqref{actiontk0}, which means that $\alpha_p=0$ and the Lagrangian reduces to the non-linear electrodynamics term $\mathscr{M}$.

 \subsection{Degenerate theories:  $\text{rk}({\EK}) =1$}
 
 We now examine the  case where $\EK$ is  still not invertible but non zero. Let us start by considering the case $\text{rk}(\EK)=1$. The theories satisfying this condition are classified in Appendix \ref{classrk1}. Given our previous result regarding quasi-linear theories, which propagates at most $3$ degrees of freedom, it is clear that we need to make additional restrictions for higher ranks to ensure as many or fewer  degrees of freedom. Thus, in the following, we are going to restrict  our study to theories involving overall only one component of the second time derivative of the gauge field.

 \subsubsection{Higher-order Maxwell theories with $\text{rk}({\EK}) =1$}
 \label{analysiscanonical}
 
As shown in \ref{classrk1}, the higher-order Maxwell Lagrangians  whose kinetic and linear parts are non-trivial but involve only one component of the second time derivative gauge field can be written in the form
 \begin{eqnarray}
 \label{rank1}
\Lag \; = \; -\frac{1}{2} (\M g^{\mu\nu}+\PP F_2^{\mu\nu}) \, \partial_\mu \N \, \partial_\nu \N \, + \, \AM \, , 
 \end{eqnarray}
 where $\AM$, $\M$, $\N$ and $\PP$ are functions of the two invariants \eqref{F2P} or, equivalently, of the two combinations
  \begin{eqnarray}
 X= \frac{1}{2}(E^2-B^2) \, , \qquad
 Y=\vec{E} \cdot \vec{B} \, ,
 \end{eqnarray}
 which we will prefer in this subsection\footnote{As showed in the appendix \ref{App:rk1},  the terms involving time derivatives of $E_i$ are of the form
$\Lag_{\dot E}=  \frac{1}{2}({\cal E}^i \dot E_i)^2 + \gamma \, {\cal E}^i \dot E_i \, , $ where ${\cal E}^i$ is a linear combination of $E^i$ and $B^i$ while $\gamma$ is a function of $X$, $Y$ and $B^2$.  
 The theory leads to two primary constraints $ C_1= E \times B \cdot \pi_E$ and  $C_2 = {\cal E}_{\perp} \cdot \pi_E$,  where $ {\cal E}_{\perp} $ belongs to the plane $(E,B)$ and normal to ${\cal E}$.  The two constraints  weakly commute and thus generate two secondary constraints. Without more constraints, the theory would contain 3 degrees of freedom.  In the rest of the section, we follow a simpler method to further study  the constraints.
}.
 For simplicity, let us start by assuming that $\PP=0$.  Instead of using the equivalent formulation \eqref{EqAction}, it is simpler here to consider the following equivalent action
 \begin{eqnarray}
 \label{equivalentrank1}
 S_{\text{eq}}[A_\mu,\phi,\lambda] =  \int d^4x \, \left[  -\frac{1}{2} \M(X,Y) \, \partial_\mu \phi \, \partial^\mu \phi +  \AM(X,Y) +  \lambda(\phi- \N(X,Y)) \right] \equiv  \int d^4x L\, ,
 \end{eqnarray}
 where $X$ and $Y$ depend on the gauge field $A_\mu$ itself. We have added the two scalars $\lambda$ and $\phi$ in order to eliminate  second derivatives
 of $A_\mu$.
 
 For the Hamiltonian analysis,  we proceed as usual and start by parametrising the phase space by the following six pairs
 of conjugate variables
 \begin{eqnarray}
 \{ A_i , \pi_i\} \, , \quad
 \{A_0 , \pi_0 \}  \, , \quad
 \{\phi,\pi_\phi \}  \, , \quad
 \{\lambda, \pi_\lambda\} \, .
 \label{phasespace}
 \end{eqnarray}
The conjugate momenta are computed from the action \eqref{equivalentrank1}, giving
 \begin{eqnarray}
 \label{momentarank1}
 \pi_\lambda = 0 \, , \quad \pi_0 = 0 \, , \quad \pi_\phi= \M \dot \phi \, , \quad 
 \pi_i = \frac{\partial \Q}{\partial \dot A^i} \equiv \Q_i= \Q_X E_i + \Q_Y B_i \, ,
 \end{eqnarray}
 with $\Q_X=\partial \Q/\partial X$ and $\Q_Y = \partial \Q/\partial Y$.
We  immediately get  two primary constraints:
\begin{eqnarray}
\pi_\lambda \simeq 0 \, , \qquad \pi_0\simeq 0 \, . 
\end{eqnarray}

If the last two expressions in (\ref{momentarank1}) can be inverted, thus allowing to  express the velocities $\dot\phi$ and $\dot A_i$ in terms of  the momenta $\pi_\phi$ and $\pi_i$, then there is no further primary constraint
and  the total Hamiltonian takes the form
\begin{eqnarray}
\label{canonicalH0}
H_0[\pi_i,\pi_\phi,A_i,\phi,\lambda] + \int d^3x \, \left(- A_0 \, \partial_i \pi^i + \mu_0 \pi_0 + \mu_\lambda \pi_\lambda \right) \, ,
\end{eqnarray}
 where $\mu_0$ and $\mu_\lambda$ are Lagrange multipliers which enforce the primary constraints. As expected, requiring the time stability  of $\pi_0 \simeq 0$ leads to the Gauss constraint $\Ga=\partial_i \pi^i \simeq 0$: these two constraints are first class and they are the generators of the gauge symmetry. The stability under time evolution of $\pi_\lambda \simeq 0$
 leads to a secondary constraint as well, that we denote $S \simeq 0$, which depends on $\lambda$ (in general) as shown in Appendix \ref{App:rk1}. These two constraints do not commute, they are second class  and  the Dirac analysis stops there. Finally, the theory propagates 3 degrees of freedom, one more than the usual two  polarisations. This extra degree of freedom is an Ostrogradski ghost. 

 \medskip

If, by contrast, the last two expressions in (\ref{momentarank1}) cannot be inverted, i.e. if they form a degenerate system,  then there exists at least one additional primary constraint and one can hope to eliminate more degrees of freedom.   This degeneracy is equivalent to  the degeneracy of the kinetic matrix, which reads
 \begin{eqnarray}
 \label{kineticbigK}
\mathbb{K} \; = \; 
\begin{pmatrix}
\dfrac{\partial^2L}{\partial \dot\phi \partial \dot \phi} & \dfrac{\partial^2  L}{\partial \dot \phi \partial E_i} \\
 \dfrac{\partial^2  L}{\partial \dot \phi \partial E^i} &  \dfrac{\partial^2 L}{\partial E^j  \partial E_i}
\end{pmatrix} \,.
 \end{eqnarray}
As shown in the appendix \ref{appendix:kinetic_rk1}, the above kinetic matrix turns out to be degenerate only if $L_X=0$.
 As a consequence, the theory \eqref{equivalentrank1} is degenerate when 
 \begin{eqnarray}
 \label{L_X=0}
 \M(X,Y)=\M(Y) \, , \qquad \N(X,Y)=\N(Y) \, , \qquad \AM(X,Y)= \AM(Y)  \, .
 \end{eqnarray}

 \medskip 
 
 When $\PP \neq 0$, the expression of  the kinetic matrix \eqref{kineticbigK} is more involved and one finds  that the
 kinetic matrix cannot be degenerate if there is a non-trivial $\PP$, as discussed in Appendix \ref{appendix:kinetic_rk1}.

 \subsubsection{Existence of an Ostrogradski ghost}
  Let us now  study the theories satisfying (\ref{L_X=0}). Because of the simple relation
 \begin{eqnarray}
 \mathscr{L} \; = \; -\frac{1}{2} \M \, \partial_\mu \N \, \partial^\mu \N  + \AM \; = \;  -\frac{1}{2} \M\N'^2\, \partial_\mu Y \, \partial^\mu Y  + \AM \, ,
 \end{eqnarray}
 we can fix $\N=Y$ without loss of generality. We prefer to work with the associated equivalent action 
\begin{eqnarray}
 S_{\text{eq}}[A_\mu,\phi,\lambda] =  \int d^4x \, \left(  -\frac{1}{2} \M(\phi) \, \partial_\mu \phi \, \partial^\mu \phi +  \AM(\phi) +  \lambda(\phi- Y) \right) \, ,
 \end{eqnarray}
which admits 5 primary constraints: the first two relations in \eqref{momentarank1} as well as
\begin{eqnarray}
\chi_i \; = \; \pi_i +\lambda B_i \,\simeq 0 , \qquad {\rm where} \quad B_i = \varepsilon_{ijk} \partial^j A^k \, .
\end{eqnarray}
The total Hamiltonian is then given by 
\begin{eqnarray}
H_{\rm tot} \; = \; H_0 + \int d^3x \, \left( \mu_0 \pi_0 + \mu_\lambda \pi_\lambda + \mu^i \chi_i \right) \,,
\end{eqnarray}
where we have added  to  the canonical Hamiltonian
\begin{eqnarray}
\label{Hamitonrank1}
H_0 \; = \; \int d^3x \, \left( \frac{\pi_\phi^2}{2 \M(\phi)} + \frac{\M(\phi)}{2} \partial_i \phi \, \partial^i \phi   - \AM(\phi) - \lambda \phi -  A_0 \partial_i \pi^i  \right) \,,
\end{eqnarray}
the primary constraints with their respective Lagrange multipliers.

The next step consists in computing the time evolution of these primary constraints. As usual,  $\pi_0 \simeq 0$ leads to the Gauss constraint and both constraints, associated with the $U(1)$ gauge symmetry, are first class. 
Computing the time evolution of the other primary constraints requires the calculation of their Poisson brackets, which are given by
\begin{eqnarray}
\{ \chi_i , \chi_j\} = \varepsilon_{ijk}\, \partial^k \lambda \, , \qquad \{ \chi_i , \pi_\lambda\} = B_i \, .
\end{eqnarray}
The  associated Dirac matrix $\Delta[\chi]$ whose coefficients are defined by the Poisson brackets, i.e.  $(\Delta[\chi])_{\mu\nu}=\{\chi_\mu, \chi_\nu\}$ (using the notation $\chi_0\equiv\pi_\lambda$ for convenience) is a $4\times 4$ antisymmetric matrix, whose determinant is
\begin{eqnarray}
\text{det}(\Delta[\chi]) = (B^i \, \partial_i \lambda)^2 \, .
\end{eqnarray}
Since $\Delta[\chi]$ is invertible generically, requiring the  time stability of these constraints determines the Lagrange multipliers $\mu_i$ and $\mu_\lambda$ and therefore does not lead to new constraints. 
 
 \medskip
 
 This completes the Dirac analysis, which has given  2 first class constraints ($\pi_0 \simeq 0$ and $\Ga \simeq 0$) and 4 second class constraints ($\chi_i \simeq 0$ and $\pi_\lambda \simeq 0)$.
 Since we started with a $12$-dimensional phase space (\ref{phasespace}), we conclude that the theory contains 2 degrees of freedom. The extra degree of freedom can be seen as an Ostrogradski ghost, since  the Hamiltonian \eqref{Hamitonrank1} is linear in $\lambda$, thus unbounded neither from above nor from below.
 
In summary, degenerate theories with $\text{rk}(\EK)=1$, and linear part in $\dot{E}$ in the same direction as the image of $\EK$, usually contains 3 degrees of freedom. With  some restrictions for the Lagrangian \eqref{L_X=0}, it is possible to eliminate one more, but it eliminates a safe degree of freedom, leaving an Ostrogradski ghost in the theory and a single  polarisation.

\subsection{Partially degenerate theories:   $\text{rk}({\EK}) =2$.}
 
 Let us finally discuss the case   $\text{rk}(\EK)=2$.  It is instructive to introduce an analogous toy model, such as 
 \begin{eqnarray}
S_2[E_i,A_i,\pi_i] = \int dt \, \, \left( \frac{1}{2} {\EK}_1  \, \dot E_1^2 +  \frac{1}{2} {\EK}_2  \, \dot E_2^2 +  C_i \, \dot E_i   - V+ \pi_i(\dot A_i - E_i)  \right) \, ,
 \end{eqnarray}
 which mimics some of the properties of the full theory. For simplicity, we assume $V$ to be quadratic in the variable $E_3$.
 We  note that the term proportional to $\dot E_3$ can be eliminated by adding to the action a boundary term $\int dt \, \dot C$ such that the function $C$ satisfies
 $ {\partial C}/{\partial E^3} = -C_3 $. We thus assume that $C_3=0$ in the following. 
 
 The part of the Lagrangian that involves time derivatives  is thus of the form
 \begin{eqnarray}
   \int dt \left[ \frac{1}{2}  \left({\EK}_1  \, \dot E_1^2 +   {\EK}_2  \, \dot E_2^2 +   {\mathscr{A}}  \, \dot A_3^2  \right) + C_1 \, \dot E_1 + C_2 \, \dot E_2 +  C_3 \, \dot A_3 \, \right],
 \end{eqnarray}
 as there are no  higher derivatives of $A_3$. If $ {\mathscr{A}}  \neq 0$, there are no constraints and the theory contains Ostrogradski ghosts. If $ {\mathscr{A}}  = 0$,
 there exists a constraint on the conjugate momentum of $A_3$, but such a constraint cannot lead to the elimination of Ostrogradski ghosts. On the contrary, it leads to the elimination
 of a regular degree of freedom. This simple analysis supports the idea that there is no healthy theory when $\text{rk}(\EK)=2$. 
 
 \medskip
 
 To illustrate this, let us consider the  simple example 
\begin{eqnarray}
 \label{rank2}
 \mathscr{L} \; = \; -\frac{1}{2} \M_1(X,Y) \, \partial_\mu X \, \partial^\mu X \, -\frac{1}{2} \M_2(X,Y) \, \partial_\mu Y \, \partial^\mu Y \, + \AM(X,Y) \, , 
 \end{eqnarray}
where $\M_1$, $\M_2$ and $\AM$ are arbitrary functions of $(X,Y)$. The kinetic matrix can be read immediately and corresponds to 
\begin{eqnarray}
{\EK}_{ij} = \frac{1}{2} \left(\M_1 E_i E_j + \M_2 B_i B_j \right) \,,
\end{eqnarray}
which is manifestly a  matrix of rank $2$. 
 
 To count and analyse its degrees of freedom, we proceed as usual and introduce the equivalent Lagrangian
 \begin{eqnarray}
 \mathscr{L}_{\rm eq}  & = & -\frac{1}{2} \M_1(\phi_1,\phi_2) \, \partial_\mu \phi_1 \, \partial^\mu \phi_1 \, -\frac{1}{2} \M_2(\phi_1,\phi_2) \, \partial_\mu \phi_2 \, \partial^\mu \phi_2 \nonumber \\
 &&+  \; \AM(\phi_1,\phi_2) \, + \lambda_1(\phi_1-X) + \lambda_2(\phi_2-Y)  \,.
 \end{eqnarray}
  The parametrisation of the phase space is similar to the previous case \eqref{phasespace} with the difference that we now have  two  scalar fields $\phi_1$ and $\phi_2$  and two Lagrange multipliers, $\lambda_1$ and $\lambda_2$, which gives a   16-dimensional phase space. One can identify three obvious primary constraints,
\begin{eqnarray}
\pi_0 \simeq 0 \, , \qquad \pi_{\lambda_1} \simeq 0 \, , \qquad  \pi_{\lambda_2} \simeq 0 \, .
\end{eqnarray}
Moreover, the
kinetic matrix is diagonal, since 
\begin{eqnarray}
\frac{\partial^2  \mathscr{L}_{\rm eq}}{\partial \dot \phi_a \, \partial \dot \phi_b} =\M_a \, \delta_{ab} \, , \qquad
\frac{\partial^2  \mathscr{L}_{\rm eq}}{\partial E_i \, \partial E_j} = - \lambda_1 \, \delta_{ij} \, , \qquad
\frac{\partial^2  \mathscr{L}_{\rm eq}}{\partial \dot \phi_a \, \partial E_i} = 0 \,,
\end{eqnarray}
and therefore  invertible if $\M_{1,2} \neq 0$, which means that  there is no further primary constraint. The rest of the analysis is straightforward:
$\pi_0 \simeq 0$, together with the Gauss constraint, suppress four phase space degrees of freedom, while the two other primary constraints  lead to two secondary constraints, the four of them being second class and eliminating four phase space degrees of freedom. We thus end up with an 8-dimensional phase space, corresponding to  4 degrees of freedom: the usual two polarisations of the electromagnetic field and two extra degrees of freedom, which behave as Ostrogradski ghosts.

\section{Conclusion}\label{SecDisc}
To summarise, we have studied the most general higher-order Maxwell action  \eqref{FlatAction} quadratic in $\partial_\alpha F_{\mu\nu}$
  and we provided strong indications that they all propagate Ostrogradski ghosts unless they reduce to a non-linear Maxwell theory (where the tensor  $\mathscr{B}=0$ vanishes identically). Even though the gauge field has several components $A_\mu$, one cannot combine them to absorb higher derivatives
  into a redefinition of variables, which is the case in higher-order scalar-tensor theories for instance. 
  This result is very similar to what has been observed for higher-order  metric theories (in 4 dimensions) which have been shown to propagate extra degrees of freedom (which are generically Ostrogradski ghosts) as well, unless they reduce to the Einstein-Hilbert action with a cosmological constant \cite{Crisostomi:2017ugk}.

 We have studied this problem according to the rank of the 3-dimensional kinetic matrix $\EK$  which is obtained by performing a 3+1 decomposition of the action \eqref{actioneqEE}. When $\text{rk}(\EK)=3$,  the theory is non-degenerate and  admits 5 degrees of freedom: the usual two polarisations of the electromagnetic field and three Ostrograski ghosts. When $\text{rk}(\EK)<3$, the theory is more complicated to analyse (at least at the Hamiltonian level) and we often introduced simpler toy models whose properties mimic some of the main properties of the full theory. We saw that the primary constraints coming from the fact that $\EK$ is degenerate are not sufficient to get rid of all the Ostrograski ghosts.

 More precisely, we saw that quasi-linear theories, i.e. imposing $\text{rk}(\EK)=0$, enables to eliminate all but one of the Ostrogradski ghost. It would be interesting to understand whether these theories can be recast as non-linear scalar-vector theories, with phantom kinetic term for the additional mode, as it is the case in the usual  Ostrogradski model. Without further assumptions, theories with $\text{rk}(\EK)>0$ would a fortiori contain more additional degrees of freedom than the previous cases. Thus, we imposed for $\text{rk}(\EK)=1$ the additional assumption that the linear term in $\dot{E}^i$ be in the same direction as the kinetic term. As we saw, this yields a  family of theories parametrised by $3$ functions of the electromagnetic invariants, with at most $3$ degrees of freedom. Imposing further that these functions depend solely on $\FFd= {}^*\!F_{\mu\nu} F^{\mu\nu}$, reduces the number of degrees of freedom to at most two, one of which, if it exists, being a ghost. Although the existence of ghosts usually imply some instabilities, it is not completely clear at this stage how such instability would manifest itself, so that a deeper understanding of degenerate higher-order Maxwell theories seems important.

 Finally, it would be interesting to extend our analysis to higher-order Yang-Mills theory in flat space, and to U(1) theories in curved space-times, i.e. to the quadratic higher-order Einstein-Maxwell theories which have been classified in \cite{Colleaux:2023cqu}. In particular, following the example of scalar-tensor theories \cite{Langlois:2015cwa}, it should be possible to systematically investigate the degeneracy conditions and degrees of freedom of these higher-order degenerate gauge theories.

\acknowledgments
We thank the Pascal Institute for the organisation of the AstroParticle Symposium 2023 where part of this research was carried out. This work was supported by the French National Research Agency (ANR) via Grant No. ANR-22-CE31-0015-01 associated with the project StronG. A.C. was partially supported by the Research Authority of the Open University of Israel (grant number 512343).

\appendix

\section{Higher Order Einstein-Maxwell theories}
\label{App:class}

In this appendix, we recall some results of \cite{Colleaux:2023cqu} 
where we considered higher-order Einstein-Maxwell theories. The corresponding  action is linear in the Riemann tensor $R_{\mu\nu\rho\sigma}$ and quadratic in the covariant derivatives
$\nabla_\alpha F_{\mu\nu}$ of the strength field, i.e.
 \begin{eqnarray}
S[g_{\mu\nu}, A_\mu] = \int d^4 x \sqrt{-g} \left(\AM + \frac{1}{4} \AR^{\mu\nu\rho\sigma} R_{\mu\nu\rho\sigma} + \AF^{\gamma\mu\nu,\delta\rho\sigma} \nabla_\gamma F_{\mu\nu}\nabla_\delta F_{\rho\sigma}  \right), \label{Action}
\end{eqnarray} 
where $\AM$ is a scalar function which depends on the two electromagnetic invariants available in four dimensions,
\begin{eqnarray}
F^2 = F^{\mu\nu}F_{\mu\nu} \,, \qquad \FFd= {}^*\!F_{\mu\nu} F^{\mu\nu} = \frac{1}{2}\varepsilon_{\mu\nu\rho\sigma} F^{\rho\sigma} F^{\mu\nu} \, ,
\label{F2P}
\end{eqnarray}
where $*$ denotes the Hodge dual defined from the Levi-Civita tensor $\varepsilon_{\mu\nu\rho\sigma}$ in four dimensions.

The tensors $\AR^{\mu\nu\rho\sigma}$ and $\AF^{\gamma\mu\nu,\delta\rho\sigma}$ are  constructed from products of the metric $g_{\mu\nu}$ and the field strength  $F_{\mu\nu}$, i.e.
\begin{eqnarray}
\AR^{\mu\nu\rho\sigma} = (F_I F_J)^{\mu\nu\rho\sigma} \, , \qquad
\AF^{\gamma\mu\nu,\delta\rho\sigma} = (F_I F_J F_K)^{\gamma\mu\nu\delta\rho\sigma} \, ,
\end{eqnarray}
where the indices  on the matrices $F_I$  are such that,
\begin{eqnarray}
F_I^{\mu\nu} = (F^I)^{\mu\nu} \, , \qquad I \in \{0,1,2,3 \} \, ,
\end{eqnarray}
{{where $F_0=g$}}.

By definition, the symmetries of $\AR^{\mu\nu\rho\sigma}$ and $\AF^{\gamma\mu\nu,\delta\rho\sigma}$ are respectively those of the Riemann tensor $R_{\mu\nu\rho\sigma}$ and of the product of the two covariant derivatives $\nabla_\gamma F_{\mu\nu}\nabla_\delta F_{\rho\sigma} $. They have been fully classified 
up to the Bianchi identity,
\begin{eqnarray}
\label{Bianchi_2}
\nabla_\mu F_{\nu\rho} + \nabla_\nu F_{\rho\mu} + \nabla_\rho F_{\mu\nu} \; = \; 0 \, , 
\end{eqnarray}
to dimensionally dependent identities (DDIs) and to boundary terms in \cite{Colleaux:2023cqu}.

The decomposition of the action \eqref{Action} 
into terms linear in the Riemann tensor and terms quadratic
in the covariant derivatives of $F_{\mu\nu}$ is not unique as one could transform terms of the first kind  into terms of second kind (and vice versa) using integrations by parts. 
We  found a basis such that $\AR^{\mu\nu\rho\sigma}$ can be decomposed according to
\begin{eqnarray}
\AR^{\mu\nu\rho\sigma} R_{\mu\nu\rho\sigma} = \sum_{n=1}^3 \aaa_n \, \mathcal{R}_n  
\end{eqnarray}
where the elementary Lagrangians  $ \mathcal{R}_n $ can be chosen as follows
\begin{eqnarray}
\label{Rnbasis}
\mathcal{R}_0\equiv R  \;,\quad
\mathcal{R}_1\equiv F^{\mu\nu} F^{\sigma\rho} R_{\mu\nu\sigma\rho} \;,\quad  \mathcal{R}_2\equiv F_{2}^{\mu\nu} R_{\mu\nu}  \;,
\end{eqnarray}
while  the functions $\AR_n$ depend on the two invariants $(F^2,\FFd)$ defined above \eqref{F2P}.  

The tensors $\AF^{\gamma\mu\nu,\delta\rho\sigma}$ are much more complicated to classify because not only we have to take into account an important number of dimensionally dependent identities but also boundary terms  in order to  carefully analyse the equivalences between such terms.  Given the basis of the Riemann type  elementary Lagrangians \eqref{Rnbasis}, we found that the most general tensor 
$\AF^{\gamma\mu\nu,\delta\rho\sigma}$ entering in the action above \eqref{Action} can be decomposed into a 18-dimensional basis,
\begin{eqnarray}
\label{Defofbbb}
 \AF^{\gamma\mu\nu,\delta\rho\sigma}  \, \nabla_\gamma F_{\mu\nu}\nabla_\delta F_{\rho\sigma}   = \sum_{n=1}^{18} \bbb_n \, \mathcal{F}_n  \, ,
\end{eqnarray}
where {{$\bbb_n$}} are still functions of the two invariants $(F^2,\FFd)$. This classification have been done in \cite{Colleaux:2023cqu} and an explicit choice for the basis elements {{$\mathcal{F}_n$}} 
has been proposed. In this basis, the elementary Lagrangians can be classified according to their weight, i.e. the number of derivatives minus the four of $\nabla F\nabla F$. There is one Lagrangian of weight 0,
\begin{eqnarray}
\label{F_1}
&&\mathcal{F}_{1}= \nabla_{\mu}F^{\mu\sigma}\nabla_{\nu}F_{\sigma}{}^{\nu} \,, \;\;  
\end{eqnarray}
7 Lagrangians of weight 2,
\begin{eqnarray}
&&\mathcal{F}_2=F_2^{\mu\nu} \nabla_{\nu}F_{\rho\sigma}\nabla^{\sigma}F_{\mu}{}^{\rho} \, ,\;\; 
\mathcal{F}_{3}=F_2^{\mu\nu}\nabla_{\sigma}F_{\gamma\nu}\nabla^{\gamma}F_{\mu}{}^\sigma  \,, \;\;
\mathcal{F}_{4}=F_2^{\mu\nu} \nabla_{\nu}F_{\mu\gamma}\nabla_{\sigma}F^{\gamma\sigma}  \,, \nonumber \\
&&\mathcal{F}_{5}=F_2^{\mu\nu} \nabla_{\sigma}F_{\mu}{}^\sigma \nabla_{\gamma}F^\gamma{}_\nu  \,, \;\; 
\mathcal{F}_6=F^{\mu\nu} F^{\rho\sigma} \nabla_{\sigma}F_{\gamma\nu}\nabla^{\gamma}F_{\mu\rho} \, , \nonumber \\
&&\mathcal{F}_{7}=F^{\mu\nu}F^{\rho\sigma} \nabla_{\nu}F_{\mu\gamma}\nabla_{\sigma}F_{\rho}{}^\gamma  \,,\;\;
\mathcal{F}_{8}=F^{\mu\nu} F^{\rho\sigma}\nabla_{\mu}F_{\nu\rho}\nabla_{\gamma}F_{\sigma}{}^\gamma  \,,\;\; 
\end{eqnarray}
1 Lagrangian of weight 3,
\begin{eqnarray}
\mathcal{F}_{9}=F^{\mu\nu}F_2^{\rho\sigma} \nabla_{\mu}F_{\nu\rho}\nabla_{\gamma}F_{\sigma}{}^\gamma \, ,
\end{eqnarray}
6 Lagrangians of weight 4,
\begin{eqnarray}
&&\mathcal{F}_{10}=F^{\mu\nu}F_3^{\rho\sigma} \nabla_{\nu}F_{\sigma\gamma}\nabla^{\gamma}F_{\mu\rho}  \,,\;\; 
\mathcal{F}_{11}=F^{\mu\nu} F_3^{\rho\sigma}\nabla_{\mu}F_{\nu\rho}\nabla_{\gamma}F_{\sigma}{}^\gamma  \,, \nonumber  \\
&&\mathcal{F}_{12}=F_2^{\mu\nu}F_2^{\rho\sigma} \nabla_{\nu}F_{\gamma\mu}\nabla_{\sigma}F_{\rho}{}^\gamma \, , \;\; 
\mathcal{F}_{13}=F_2^{\mu\nu}F_2^{\rho\sigma} \nabla_{\mu}F_{\nu\rho}\nabla_{\gamma}F_{\sigma}{}^\gamma  \,,  \nonumber  \\
&&\mathcal{F}_{14}=F^{\mu\nu} F^{\rho\sigma}F_2^{\gamma\delta} \nabla_{\mu}F_{\nu\rho}\nabla_{\delta}F_{\sigma\gamma}  \,,\;\; 
\mathcal{F}_{15}=F^{\mu\nu}F^{\rho\sigma} F_2^{\gamma\delta}\nabla_{\nu}F_{\mu\gamma}\nabla_{\sigma}F_{\rho\delta} \, , \,\;\;
\end{eqnarray}
2 Lagrangians of weight 5,
\begin{eqnarray}
\mathcal{F}_{16}=F_2^{\mu\nu}F_3^{\rho\sigma} \nabla_{\mu}F_{\nu\rho}\nabla_{\gamma}F_{\sigma}{}^{\gamma}  \,, \;\;
\mathcal{F}_{17}=F^{\mu\nu}F_2^{\rho\sigma} F_2^{\gamma\delta}\nabla_{\nu}F_{\mu\rho}\nabla_{\delta}F_{\sigma\gamma}  \,,   
\end{eqnarray}
and one last Lagrangian of weight 6,
\begin{eqnarray}
\label{F_18}
\mathcal{F}_{18}=F^{\mu\nu}F_2^{\rho\sigma} F_3^{\gamma\delta}\nabla_{\nu}F_{\mu\gamma}\nabla_{\sigma}F_{\rho\delta} \, . 
\end{eqnarray}
When the space-time is flat, the Riemann curvature vanishes and  the tensors $\AR^{\mu\nu\rho\sigma}$ thus become irrelevant. Furthermore, the
term $\mathcal{F}_2$ in the previous basis becomes redundant  because it can be shown to be equivalent  (up to boundary terms) to a combination of the other elementary 
Lagrangians and the term   $F^{\mu\nu} F_3^{\rho\sigma} R_{\mu\nu\rho\sigma}$ which obviously vanishes when the metric is flat, as it can be seen from Eq. (6.8) of \cite{Colleaux:2023cqu} .

\section{Kinetic Matrix in the flat case}
\label{App}

One can explicitly compute the  kinetic matrix \eqref{Gigen} which  can be shown to decompose as follows,
\begin{eqnarray}
\EK^{ij} =\EK_1 B^i B^j +\EK_2 E^i E^j+ \EK_3 E^{(i}B^{j)} +\EK_4 h^{ij} +\EK_5 B^{(i} \Pi^{j)}  +\EK_6 E^{(i} \Pi^{j)}  \,,\label{KinMatrixEBPi}
\end{eqnarray}
where we have defined  the Poynting vector $\Pi_i= \left( E \times B\right)_i  = \varepsilon_{ijk} E^jB^k $. The coefficients $\EK_I$ depend on the functions $\bbb_n$ introduced  in the Lagrangian \eqref{Defofbbb}, according to
\begin{eqnarray}
\begin{split}
\EK_1&= \frac{1}{2}  \left( F^2 - 2 B^2 \right)\bbb_{12} - \left( \bbb_3 + \bbb_{5} \right) \, ,\\
\EK_2&=- \left(\bbb_{3}+\bbb_{5}-\bbb_{6}+\bbb_{7}+\bbb_{8} \right)+ \frac{F^2}{4} \left(2 \left(-\bbb_{10} +\bbb_{11}+\bbb_{13}+\bbb_{14}+\bbb_{15}\right)-F^2 \bbb_{18} \right)  \, , \\
&\phantom{=}- B^2 \left(\bbb_{12}+\bbb_{14}+\bbb_{15}-\frac{F^2}{2} \bbb_{18} \right) \, , \\
\EK_3&=E.B \left(\frac{1}{2}\left(\bbb_{10}-\bbb_{11}-B^2 \bbb_{18}  \right) +\bbb_{12}-\bbb_{13}+ \frac{F^2}{4} \bbb_{18}  \right) \, , \\
\EK_4&=-\bbb_{1} + B^2 \left(\bbb_{3}-\bbb_{4}+\bbb_{5}+\frac{F^2}{2} \bbb_{12}\right) +\frac{F^2}{4} \left(2 \bbb_{4} + F^2 \bbb_{12} \right)+\left( E.B\right)^2\left( -\bbb_{12} +\bbb_{13}\right) \, , \\
\EK_5&= \frac{1}{2} E.B \, \bbb_{16} \, , \\
\EK_6&=\frac{1}{4} \left( 2\bbb_{9}-F^2 \bbb_{16} \right) \, .
\end{split}
\end{eqnarray}
Notice that we have used  $\FFd=-4 E\cdot B$ but we have kept $F^2$.  Let us also remark that we could have replaced  $h^{ij}$ by its expression in terms of  $\Pi^i \Pi^j $ in order to express $\EK^{ij}$ in the six-dimensional basis of $3\times 3$ symmetric matrices associated with the vectorial basis $\left\{E,B,\Pi \right\}$.

\section{More on the analysis of theories with $\text{rk}(\EK)=1$}
\label{App:rk1}

In this appendix, we study in more details theories such that $\text{rk}(\EK)=1$. We first propose a classification of these theories (with supplementary conditions that we are going to describe below) and then  give more details on their constraint analysis.

\subsection{Classification of theories with $\text{rk}(\EK)=1$}\label{classrk1}

Requiring that the kinetic matrix \eqref{KinMatrixEBPi} has rank 1 amounts to find two vectors of the form 
\be
\label{condonrank1}
V =v_1 E  + v_2 B + v_3 E \!\times \!B
\ee
such that $\EK^{ij} V_j=0$. This results in the following system of equations, 
\begin{eqnarray}
\begin{pmatrix}
\EK_4+ E.B \EK_3 + E^2 \EK_2   &  E.B \EK_2 +B^2 \EK_3 & \left(B^2 E^2 -\left( E.B\right)^2 \right)\EK_6 \\
E.B \EK_1 + E^2 \EK_3 & \EK_4 + B^2 \EK_1 + E.B \EK_3 &   \left(B^2 E^2 -\left( E.B\right)^2 \right)\EK_5  \\
 E.B \EK_5+E^2 \EK_6  & B^2 \EK_5 + E.B \EK_6 & \EK_4 
 \end{pmatrix}
 \begin{pmatrix}
 v_1 \\
 v_2 \\
 v_3
 \end{pmatrix}
 = 0 \,.
 \end{eqnarray}
It is convenient to write the previous matrix in the form $\left( \vec{Z}_1, \vec{Z}_2, \vec{Z}_3 \right) $
where the vectors $\vec{Z}_i$ are obviously proportional to each other as the matrix is of rank 1.  
Thus,  all the determinants of the minors of the matrix should vanish. For instance
\begin{eqnarray}
\det  \begin{pmatrix}
 W_{12} & W_{32} \\
 W_{13} &  W_{33}
 \end{pmatrix} = E^2  \EK_3  \EK_4 -  \frac{E.B}{B^2}  \left( E.B  \EK_3 +  \EK_4 \right)  \EK_4 +\left( \left( E.B \right)^2 - B^2 E^2  \right)^2  \frac{\EK_5  \EK_6}{B^2} =0\,,
\end{eqnarray}
where we introduced the following vectors for simplicity,
\begin{eqnarray}
\vec{W}_1 =  \vec{Z}_1 - \frac{E.B}{B^2}  \vec{Z}_2 \,,\;\;\;\;\vec{W}_2 =  \vec{Z}_1 - \frac{E^2}{E.B}  \vec{Z}_2  \,,\;\;\;\; \vec{W}_3 =\vec{Z}_3 \,.
\end{eqnarray}
All these equations are viewed as polynomials of $B^2$ whose ``coefficients'' are in fact functions of $X$ and $E.B$ which must vanish. 

Imposing all these conditions fixes 7 functions among the 17 available ones. Hence, the resulting theory can be written as a sum of the 7 quasi-linear Lagrangians (of vanishing rank) and a 3-functions-family of theories with non-vanishing kinetic term, i.e.
\begin{eqnarray}
S[A_\mu]  =  \int d^4 x  \, \big( \AM +  \sum_{p=1}^7 \alpha_p \, \mathscr{L}_p  +  \mathscr{L} \big) \,,
\end{eqnarray}
where the quasi-linear theories are defined by \eqref{QLT2} while the part with a non-vanishing kinetic terms  can be written in the form, 
\begin{eqnarray}
\mathscr{L}=  \xi_1^2 \left( \gamma_1\mathcal{F}_{8}+ \gamma_2  \mathcal{F}_{15} \right) -4 \xi_1 \xi_2 \left( \gamma_1 \mathcal{F}_{11} + \gamma_2 \mathcal{F}_{18}\right)+ \xi_2^2 \left( \gamma_1 \mathscr{F}_1 + \gamma_2\mathscr{F}_2  \right)
\end{eqnarray}
with 
\begin{eqnarray}
\begin{split}
\mathscr{F}_1 &= 2 \left(E.B\right)^2\left( F^2 \mathcal{F}_{1}+2\left( \mathcal{F}_{4}+\mathcal{F}_{5}-\mathcal{F}_{8}\right)- F^2 \left( F^2 \mathcal{F}_{8}+4 \mathcal{F}_{11}\right)  \right)  \,,\\
 \mathscr{F}_2 &=-4 \left(E.B\right)^4 \mathcal{F}_{1} + 2 \left(E.B\right)^2 \left(  F^2 \left( \mathcal{F}_{4}+\mathcal{F}_{8} \right) + 2 \left(2 \mathcal{F}_{11}+\mathcal{F}_{12}-\mathcal{F}_{15}\right)\right) - F^2 \left(F^2 \mathcal{F}_{15}+ 4 \mathcal{F}_{18}\right) \,,
\end{split}
\end{eqnarray}
where $\gamma_m$ and $\xi_m$, with $m=1,2$, are functions of the electromagnetic (EM) invariant. Notice that one out of the two functions $\gamma_1$ and $\gamma_2$ is redundant. 

It is straightforward to obtain the kinetic term of this model, 
\begin{eqnarray}
\mathscr{L} =  - \left(  \gamma_1 + \gamma_2 E^2\right) \left(  \left( \xi_1 + \xi_2 F_2 \right)E^i\dot{E}_i -2 \xi_2 E.B B^i  \dot{E}_i\right)^2 + \mathscr{L}_{lin}+ \mathscr{L}_{pot} \label{LkinRank1}
\end{eqnarray}
where $\mathscr{L}_{lin}$ and $\mathscr{L}_{pot}$ are respectively linear in the  velocities and without velocities. In order to guarantee the absence of Ostrogradski ghosts associated with the two components of the velocity which are absent from the kinetic term, we impose these to be absent from the linear term as well, which implies that $ \mathscr{L}_{lin}$
is of the form
\begin{eqnarray}
\label{LlinRank1}
 \mathscr{L}_{lin} \propto \dot{E}_i  \left(  \left( \xi_1 + \xi_2 F_2 \right)E^i -2 \xi_2 E.B B^i  \right) \, .
\end{eqnarray}
Let us  present an explicit 3-parameter family of theories which satisfy  the required properties\footnote{It is not simple to classify theories with this property. One of the reasons is  that there are boundary terms relating this linear term to the potential and we should take all of these into account. For example, it is clear that the velocity in $\dot{E} . \partial \times B$ can be traded for spatial derivatives of the electric field using the Bianchi identity. Moreover, as they involve a time derivative, these boundary terms must be covariant. As shown in \cite{Colleaux:2023cqu}, there are 5 independent such terms in four-dimensions. It is quite cumbersome and not particularly illuminating to obtain the conditions that the theories must satisfy from this perspective.
}, 
\begin{eqnarray}
\mathscr{L}  =\frac{1}{16} \left( \xi_1 \partial_\mu F^2 + \xi_2 \partial_\mu F_4 \right) \left( \gamma_1  g^{\mu\nu} + \gamma_2 F_2^{\mu\nu} \right) \left( \xi_1 \partial_\nu F^2 + \xi_2 \partial_\nu F_4 \right) 
\end{eqnarray}
where
\begin{eqnarray}
F_4 = \frac{1}{4} \left(\FFd^2 + 2\left( F^2\right)^2 \right)\,.
 \end{eqnarray}
We see that both conditions \eqref{LkinRank1} and \eqref{LlinRank1} are satisfied. As a consequence, we obtain the most general theory\footnote{The independence of these terms implies that this theory is the one we look for. Indeed, this independence can be understood from the following argument : the dimensionally dependent identities in four dimensions impose that there are two independent EM scalars, so that $F^2$ and $F_4$, and thus also $\partial_\mu F^2$ and $\partial_\mu F_4$ can be considered independent. Given the high symmetry in $\mathcal{L}$, there are no more dimensional identities that could be used to reduce these 3 terms. It is clear that instead of choosing a basis of EM invariants $\{F^2,F_4\}$, it is possible to consider alternatively $\{X,Y\}$, where $X$ and $Y$ are two different functions of the invariants.} which can be equivalently reformulated as \eqref{rank1},
 \begin{eqnarray}
\Lag \; = \; -\frac{1}{2} (\M g^{\mu\nu}+\PP F_2^{\mu\nu}) \, \partial_\mu \N \, \partial_\nu \N \, + \, \AM \, , 
 \end{eqnarray}
 where $\AM$, $\M$, $\N$ and $\PP$ are functions of the two invariants \eqref{F2P} or, equivalently, of the two combinations
  \begin{eqnarray}
 X= \frac{1}{2}(E^2-B^2) \, , \qquad
 Y=\vec{E} \cdot \vec{B} \, ,
 \end{eqnarray}

\medskip

Finally, notice that many other theories admitting a rank $1$ exist, but they involve higher powers of $\partial F$. For instance, it is possible to generate a rank $1$ quartic theory applying the following non-invertible transformations to the Maxwell action, $A_\mu \longrightarrow A_\mu +\FFd  \partial_\nu F^2$, while applying it on the other invariant $\FFd$ yields a quadratic quasi-linear theory classified by \eqref{QLT2}. More generally, it is clear that the square of any theory with rank $0$ is a theory with rank $1$.

\subsection{Constraint analysis} 
We consider theories whose dynamics is governed by the action \eqref{rank1}, or equivalently \eqref{equivalentrank1}, where we fix $\PP=0$ for simplicity,
\begin{eqnarray}
S_{\text{eq}}[A_\mu,\phi,\lambda] =  \int d^4x \, \left[  -\frac{1}{2} \M(X,Y) \, \partial_\mu \phi \, \partial^\mu \phi +  \AM(X,Y) +  \lambda(\phi- \N(X,Y)) \right] \, .
\end{eqnarray}
All the notations have already been introduced in the paper. We have started the analysis in section \ref{analysiscanonical} but we have not given neither the form of the canonical Hamiltonian $H_0$ in
\eqref{canonicalH0} nor the expression of the secondary constraint $S \simeq 0$ which comes when one study the stability  of the primary constraint $\pi_\lambda \simeq 0$.  Here, we fill this gap.

\subsubsection{Completing the canonical analysis}
In order to compute the canonical Hamiltonian $H_0$, we assume that the theory is not degenerate in the photon sector which means that the relation 
\begin{eqnarray}
\label{PiEE}
\pi_i = \Q_X E_i + \Q_Y B_i \, ,
\end{eqnarray}
is invertible and then one can express the ``velocities'' $E_i$ in terms of the momenta $\pi_i$. In general, such an inversion cannot be done  explicitely  and one obtains an implicit expression for $E_i$,
\begin{eqnarray}
\label{EpiU}
E_i = U_i (\pi_j,\pi_\phi) \, .
\end{eqnarray}
At this stage, one can compute the canonical Hamiltonian which is given by
\begin{eqnarray}
H_0 = \int d^3x \, \left[ \pi^i U_i + \frac{\pi_\phi^2}{2 \M} + \frac{1}{2} \M (\partial_i \phi)^2 - {\AM} - \lambda (\phi-\N)\right]\, .
\end{eqnarray}
Notice that the Hamiltonian should be defined with a canonical term $\pi^i \dot A_i$ instead of  $\pi^i U_i = \pi^i( \dot A_i - \partial_i A_0)$. However, the term $\pi^i \partial_i A_0$ leads (after an integration by parts) to the Gauss constraint which has already been taken into account in the expression of the total Hamiltonian \eqref{canonicalH0}. 

Now   $\M$, $\N$ and $\AM$ (and also $U_i$) are viewed as functions of the momenta and they also depend on $\lambda$ because the function $\Q$ in \eqref{EpiU} depends on $\lambda$.  Therefore, requiring the stability under
time evolution of the primary constraint $\pi_\lambda \simeq 0$ leads to a secondary constraint $S \simeq 0$ where
\begin{eqnarray}
S = \frac{\delta H_0}{\delta \lambda} = \pi^i \partial_\lambda U_i + \left(- \frac{\pi_\phi^2}{2\M^2}  + \frac{1}{2}  (\partial_i \phi)^2 \right) \partial_\lambda \M - \partial_\lambda \AM + \lambda \partial_\lambda \N + \N-\phi \, .
\end{eqnarray}
Using the property
\begin{eqnarray}
\partial_\lambda \M = \frac{\partial \M}{\partial E_i} \partial_\lambda U_i = (\M_X E^i + \M_Y B^i) \partial_\lambda U_i \, ,
\end{eqnarray}
that holds also for the  functions $\N$ and $\AM$, we immediately see that
\begin{eqnarray}
S = \left( \pi^i - Q_X E^i - \Q_Y B^i\right) \partial_\lambda U_i + \N - \phi = \N - \phi \, , 
\end{eqnarray}
where we have used the expression of $\pi_i$ \eqref{PiEE}. As expected, we recover that requiring the stability under time evolution of $\pi_\lambda \simeq 0$ leads to 
the constraint $\N-\phi \simeq 0$ which is the Euler-Lagrangian equation for $\lambda$. The main difference is that now $\N$ is viewed as a function of the momenta and no more on
the velocities. When one inverts \eqref{PiEE}, one gets \eqref{EpiU} where $U_i$ depends non trivially on $\lambda$. As a consequence, $S$ depends non trivially on $\lambda$ and
then the Dirac analysis closes here, as we claim in  in section \ref{analysiscanonical}.

\subsubsection{An alternative analysis}
The previous class of theories with $\text{rk}(\EK)=1$ can be alternatively described by an action of the form
\begin{eqnarray}
 \int d^4x \, \left[  -\frac{1}{2} \, \left( \M_1(X,Y) \partial_\mu X + \M_2(X,Y) \partial_\mu Y \right) \left( \M_1(X,Y) \partial^\mu X + \M_2(X,Y) \partial^\mu Y \right)  +  \AM(X,Y) \right]
\end{eqnarray}
where the functions $\M_1$ and $\M_2$ are related to the functions $\M$ and $\N$ entering in \eqref{rank1} by
\begin{eqnarray}
\M_1^2= \M \, \N_X^2 \, , \qquad \M_2^2= \M \N_Y^2 \, ,
\end{eqnarray}
i.e. we have assumed that $\M>0$ (which is natural if one wants to avoid ghosts).

 This new action is equivalent to 
\begin{eqnarray}
 \int d^4x \, \left[  -\frac{1}{2} \, \left( \M_1 \partial_\mu \phi_1 + \M_2 \partial_\mu \phi_2 \right) \left( \M_1 \partial^\mu \phi_1 + \M_2 \partial^\mu \phi_2 \right)  +  \AM +  \lambda_1(\phi_1- X) + \lambda_2 (\phi_2 - Y) \right] 
\end{eqnarray}
where now the $\M_1$, $\M_2$ and $\AM$ are viewed as function of the scalar fields $\phi_1$ and $\phi_2$. 

Its analysis is similar to the previous one. We start with a parametrisation of the phase space in terms of the following pairs of conjugate variables
\begin{eqnarray}
 \{ A_i , \pi_j\}= \delta_{ij} \, , \quad
 \{A_0 , \pi_0 \} = 1 \, , \quad
 \{\phi_a,\pi_{\phi_a} \} = 1 \, , \quad
 \{\lambda_a, \pi_{\lambda_a}\} = 1 \, ,
 \label{phasespace2}
 \end{eqnarray}
 where $a \in \{1,2\}$.  The only non-trivial momenta are given by,
 \begin{eqnarray}
 \pi_i = -\lambda_1 E_i - \lambda_2 B_i \, , \qquad \pi_{\phi_a} = \M_a (\M_1 \dot \phi_1 + \M_2 \dot \phi_2) \, .
 \end{eqnarray}
Therefore the theory admits the following primary constraints,
\begin{eqnarray}
\pi_0 \simeq 0 \, , \qquad
\pi_{\lambda_a} \simeq 0 \, , \qquad
\chi=\M_2 \pi_{\phi_1} - \M_1 \pi_{\phi_2} \simeq 0 \, .
\end{eqnarray}
At this stage, one can compute the total Hamiltonian  which takes a form similar to \eqref{canonicalH0},
\begin{eqnarray}
H_0 + \int d^3x \, \left(- A_0 \, \partial_i \pi^i + \mu_0 \pi_0 + \mu_{\lambda_a} \pi_{\lambda_a} + \mu_\chi \chi \right) \, ,
\end{eqnarray}
where the Lagrange multipliers  $ \mu_0$, $\mu_{\lambda_a}$ and $\mu_\chi$ enforce the primary constraints while the canonical Hamiltonian is given by
\begin{eqnarray}
H_0 = \int d^3x \, \left[ 
\frac{\pi_{\phi_1}^2}{2 \M_1^2} + \frac{1}{2} \left( \M_1 \partial_i \phi_1 + \M_2 \partial_i \phi_2 \right)^2 - \frac{\pi^2 + 2 \lambda_2 \pi \!\cdot \!B + (\lambda_1^2 + \lambda_2^2)B^2}{2\lambda_1}   - \lambda_1 \phi_1 - \lambda_2 \phi_2  \right] .\nonumber
\end{eqnarray}
Requiring the stability of $\pi_0$ leads, as usual, to the Gauss $G \simeq 0$ constraint and both are first class. 

Following the same method as in the previous subsection, we can show that studying the stability under time evolution of the two constraints $\pi_{\lambda_a} \simeq 0$ leads to 
the expected two secondary constraints $S_a \simeq 0$ with,
\begin{eqnarray}
S_1&=& \phi_1 - X = \phi_1 - \frac{\pi^2 + 2\lambda_2 \pi \!\cdot \!B +(\lambda_2^2 - \lambda_1^2)B^2}{2 \lambda_1^2} \, , \\
S_2&=&\phi_2-Y= \phi_2 + \frac{\pi \!\cdot \!B + \lambda_2 B^2}{\lambda_1} \, .
\end{eqnarray}
Taken together, $\pi_{\lambda_a}$ and $S_a$ form a set of second class constraints as their corresponding Dirac matrix is invertible. In other words, we can use these four constraints to eliminate
$\lambda_a$ and their momenta by expressing them in terms of the remaining phase space variables.

Finally, we are left with the constraint $\chi \simeq 0$ whose stability leads to a secondary constraint $\psi \simeq 0$ and, in general, the Dirac analysis stops here. Hence, the theory admits
2 first class constraints and three pairs of second class constraints. As we started with 8 pairs of variables in the phase space, we conclude that the theory admits $3\times 2$ physical degrees of freedom
in the phase space. This is consistent of the result of the analysis of the theory when formulated in terms of the action \eqref{equivalentrank1}.

\subsection{Degeneracy of the kinetic matrix in the case $\text{rk}(\EK)=1$}
\label{appendix:kinetic_rk1}

The kinetic matrix associated with the rank $1$ theory \eqref{rank1} is given by 
 \begin{eqnarray}
\mathbb{K} \; = \; 
\begin{pmatrix}
\dfrac{\partial^2L}{\partial \dot\phi \partial \dot \phi} & \dfrac{\partial^2  L}{\partial \dot \phi \partial E_i} \\
 \dfrac{\partial^2  L}{\partial \dot \phi \partial E^i} &  \dfrac{\partial^2 L}{\partial E^j  \partial E_i}
\end{pmatrix} = 
 \begin{pmatrix}
 \M & \dot \phi \M^i\\
 \dot \phi \M_j & \Q^i_j 
 \end{pmatrix}
 \end{eqnarray}
where
 \begin{eqnarray}
 \M_i &= & \M_X E_i + \M_Y B_i \, , \\
 \Q_{ij} &= & \Q_X \delta_{ij} + \Q_{XX} E_i E_j + \Q_{YY} B_i B_j + \Q_{XY} (E_i B_j + E_j B_i) \,,
 \end{eqnarray}
 and indices are lowered and raised with the  Kronecker delta $\delta_i^j$.

 The kinetic matrix is degenerate if it admits at least one null direction ${}^{t}(v_0,V_i)$. Because of the covariance, we know 
 that the ``spatial'' component of $V$ can be decomposed according to \eqref{condonrank1}.
 Then, the condition ${\mathbb K} V=0$ decomposes into a single equation
 for $v_3$
 \begin{eqnarray}
 \Q_X v_3 \; = \; 0 \, , 
 \end{eqnarray}
 and a more involved system of equations for $(v_0,v_1,v_2)$,
 \begin{eqnarray}
 \label{matrixdeg_2}
 \begin{pmatrix}
 \M & \dot \phi (2X \M_X + Y \M_Y + B^2 \M_X) & \dot \phi(Y \M_X + B^2 \M_Y) \\
 \dot \phi \M_X & Q_X + 2X Q_{XX} + Y \Q_{XY} + B^2 \Q_{XX} & Y \Q_{XX} + B^2 \Q_{XY} \\
 \dot \phi \M_Y & Y \Q_{YY} + 2X \Q_{XY} + B^2 \Q_{XY} & \Q_X + Y \Q_{XY} + B^2 \Q_{YY}
 \end{pmatrix}
 \begin{pmatrix}
 v_0 \\
 v_1 \\
 v_2
 \end{pmatrix}
 = 0 \,,
 \end{eqnarray}
 where we have used the relations $E^2=2X+B^2$ to eliminate the terms $E^2$ from these equations. 
 
 If $\Q_X=0$, the kinematic matrix is clearly degenerate and admits at least one nul direction defined by $v_0=0$ and $V= E \!\times \!B$.
 
 Another possibility is that the sub-system \eqref{matrixdeg_2} is itself degenerate and 
 therefore admits a non trivial solution. This happens if the associated three dimensional matrix \eqref{matrixdeg_2} has a vanishing determinant which can be expanded as follows
 \begin{eqnarray}
 \text{Det} \, = \, (C_0^0 + C_0^2 B^2 + C_0^4 B^4) + \dot \phi^2  (C_2^0 + C_2^2 B^2 + C_2^4 B^4)  \, ,
 \end{eqnarray}
 where the functions $C_p^n(X,Y)$ can be expressed in terms of $\M$, $\Q$ and their derivatives with respect to $X$ and $Y$. As a consequence,
 requiring the degeneracy leads to the 6 relations
 \begin{eqnarray}
 C_p^n(X,Y) \; = \; 0 \; , \qquad p \in \{0,2\} \, , \quad  n \in \{0,2,4\} \, ,
 \end{eqnarray}  
 which is a system of  non linear partial differential equations. Interestingly, the relation $C_0^4=0$ is particularly simple and reduces to the Monge-Ampère equation for the function 
 $\Q$,
 \begin{eqnarray}
 \label{MAequation}
 \Q_{XX} \Q_{YY} - \Q_{XY}^2 = 0 \, ,
 \end{eqnarray}
 assuming that $\M \neq 0$. This equation is reminiscent from the fact that the kinetic matrix is nothing but a Hessian matrix. When one substitutes this relation into $C_0^2$, one obtains
 \begin{eqnarray}
 \Q_X \, (\Q_{XX} + \Q_{YY}) \; = \; 0 \, ,
 \end{eqnarray}
 which leads to two branches of solutions a priori. If we assume that $\Q_X=0$, then all the remaining conditions $C_p^n(X,Y)=0$ can be shown to be trivially satisfied.
 In the second branch, the condition $\Q_{XX} + \Q_{YY}=0$ together with the Monge-Ampère equation \eqref{MAequation} leads to the fact that he Hessian matrix of the function $\Q$ vanishes identically, i.e.
 $\Q_{XX}=\Q_{YY}=\Q_{XY}=0$. When one substitutes these conditions into  $C_0^0(X,Y)=0$, one gets $\Q_X=0$ and all the remaining conditions $C_p^n(X,Y)=0$ are  satisfied as in the first branch. 

\medskip

When $\PP \neq 0$, the expression of  the kinetic matrix \eqref{kineticbigK} is more involved. Hence the conditions for it to be degenerate are also more complicated but it is possible to check that the 
 kinetic matrix cannot be degenerate if there is a non-trivial $\PP$.

 Let us give an argument to explain this result. First we compute the momenta
 which are now given by,
 \begin{eqnarray}
 \pi_\phi & = & (M + 2X \PP + B^2 \PP) \dot \phi - \PP (E\cdot B \! \times \! \partial \phi ) \, , \\
 \pi_i & = & \Q_i -  \left[ \PP_i (E \cdot B \! \times \! \partial \phi) - \PP(B \! \times \! \partial \phi)_i  \right] \dot \phi  - \frac{1}{2} \PP_i (E \cdot \partial \phi)^2 - \PP(E \cdot \partial \phi) \, \partial_i \phi \, .
   \end{eqnarray}
 If we proceed as we did above, we express $\dot \phi$ in terms of $\pi_\phi$ and we substitute its expression into $\pi_i$. Then, we notice that the momenta $\pi_i$ involve, the following three 
 components of $E$ 
 \begin{eqnarray}
E \cdot B \, , \qquad  E \cdot B \! \times \! \partial \phi  \, , \qquad E \cdot \partial \phi \, .
 \end{eqnarray}
 Therefore, even if $\PP$ and $\Q$ depends only on $Y$, we could in principle express three independent components of $E$ in terms of the momenta which would make the system
 invertible and thus the kinetic matrix non-degenerate.

\bibliographystyle{utphys}
\bibliography{Biblio_Einstein_Maxwell_2}

\end{document}